\documentclass[3p,preprint]{elsarticle}

\journal{Theoretical Computer Science}

\usepackage{amssymb}

\usepackage[figuresright]{rotating}

\newtheorem{theorem}{Theorem}
\newtheorem{lemma}[theorem]{Lemma}
\newtheorem{proposition}[theorem]{Proposition}
\newtheorem{corollary}[theorem]{Corollary}
\newtheorem{definition}[theorem]{Definition}
\newtheorem{claim}{Claim}
\newtheorem{example}{Example}
\newtheorem{remark}{Remark}

\newproof{proof}{Proof}
\newproof{proofsketch}{Proof Sketch}

\usepackage[utf8]{inputenc}
\usepackage{complexity}
\usepackage[algosection]{algorithm2e}
\usepackage{endnotes}
\usepackage{listings}
\usepackage{mathtools}
\usepackage{verbatim}
\usepackage{mathrsfs}
\usepackage{longtable}
\usepackage{enumitem}
\usepackage{etoolbox}
\usepackage{float}
\usepackage{color}
\usepackage{hyperref}

\usepackage{linenofb}

\usepackage{complexity}
\usepackage{comment}
\usepackage{longtable}

\newenvironment{code}
{\par\runninglinenumbers
\modulolinenumbers[1]%
\linenumbersep-1em
\def\linenumberfont
{\normalfont\itshape}}

\renewcommand{\t}{\hspace*{3mm}}

\newcommand{\shuffle}{\hspace{1mm}{\mathbin{\mathchoice
{\rule{.3pt}{1ex}\rule{.3em}{.3pt}\rule{.3pt}{1ex}
\rule{.3em}{.3pt}\rule{.3pt}{1ex}}
{\rule{.3pt}{1ex}\rule{.3em}{.3pt}\rule{.3pt}{1ex}
\rule{.3em}{.3pt}\rule{.3pt}{1ex}}
{\rule{.2pt}{.7ex}\rule{.2em}{.2pt}\rule{.2pt}{.7ex}
\rule{.2em}{.2pt}\rule{.2pt}{.7ex}}
{\rule{.3pt}{1ex}\rule{.3em}{.3pt}\rule{.3pt}{1ex}
\rule{.3em}{.3pt}\rule{.3pt}{1ex}}\mkern2mu}}\hspace{1mm}}

\usepackage{mathtools}

\newcommand*{\DEBUG}{}%

\ifdefined\DEBUG
\newcommand{\fixme}[1]{{\textcolor{red}{\bf{\textsf{FIXME: #1}}}}}
\newcommand{\bug}[1]{{\textcolor{blue}{\bf{\textsf{BUG: #1}}}}}
\newcommand{\idea}[1]{{\textcolor{blue}{\bf{\textsf{IDEA: #1}}}}}

\newcommand{\TODO}[1]{{\textcolor{red}{\bf{\textsf{ 
TODO: #1
}}}}}
\else
\newcommand{\fixme}[1]{}
\newcommand{\bug}[1]{}
\newcommand{\TODO}[1]{}
\newcommand{\idea}[1]{}
\fi

\newclass{\COMSLIP}{COM\mbox{-}SLIP}
\newclass{\COMSLIPCUP}{COM\mbox{-}SLIP^{\cup}}

\newclass{\DCM}{DCM}
\newclass{\eDCM}{eDCM}
\newclass{\eNPDA}{eNPDA}
\newclass{\DPDA}{DPDA}
\newclass{\RDPDA}{RDPDA}
\newclass{\PDA}{PDA}
\newclass{\DCMNE}{DCM_{NE}}
\newclass{\TwoDCM}{2DCM}
\newclass{\NCM}{NCM}
\newclass{\eNCM}{eNCM}
\newclass{\eNQA}{eNQA}
\newclass{\eNSA}{eNSA}
\newclass{\eNPCM}{eNPCM}
\newclass{\eNQCM}{eNQCM}
\newclass{\eNSCM}{eNSCM}
\newclass{\DPCM}{DPCM}
\newclass{\NPCM}{NPCM}
\newclass{\NQCM}{NQCM}
\newclass{\NSCM}{NSCM}
\newclass{\NPDA}{NPDA}
\newclass{\TRE}{TRE}
\newclass{\NFA}{NFA}
\newclass{\DFA}{DFA}
\newclass{\NCA}{NCA}
\newclass{\DCA}{DCA}
\newclass{\DTM}{DTM}
\newclass{\NTM}{NTM}
\newclass{\DLOG}{DLOG}
\newclass{\CFG}{CFG}
\newclass{\ETOL}{ET0L}
\newclass{\EDTOL}{EDT0L}
\newclass{\CFP}{CFP}
\newclass{\ORDER}{O}
\newclass{\MATRIX}{M}
\newclass{\BD}{BD}
\newclass{\LB}{LB}
\newclass{\ALL}{ALL}
\newclass{\decLBD}{decLBD}
\newclass{\StLB}{StLB}
\newclass{\SBD}{SBD}
\newclass{\TCA}{TCA}
\newclass{\TMRB}{TMRB}
\newclass{\TMFC}{TMFC}

\newsavebox{\spacebox}
\begin{lrbox}{\spacebox}
\verb*! !
\end{lrbox}
\newcommand{\LL}{{\cal L}}

\DeclareMathOperator{\alp}{alph}
\DeclareMathOperator{\fulltrio}{\hat{\cal T}}
\DeclareMathOperator{\fullsemiafl}{\hat{\cal S}}

\newcommand\bd{^{{\rm bd}}}
\newcommand\fin{_{{\rm fin}}}

\begin{document}

\begin{frontmatter}




\title{On Families of Full Trios Containing Counter Machine Languages\tnoteref{t1}\tnoteref{t2}}

\tnotetext[t1]{\textcopyright 2022. This manuscript version is made available under the CC-BY-NC-ND 4.0 license \url{http://creativecommons.org/licenses/by-nc-nd/4.0/} The manuscript is published in O.H. Ibarra, I. McQuillan. On Families of Full Trios Containing Counter Machine Languages. {\it Theoretical Computer Science} 799, 71--93 (2019).}

\tnotetext[t2]{A preliminary conference version of this paper appeared in \cite{DLT2016}. This extended version includes all proofs and many new results including all of Sections \ref{decproperties} and \ref{weakproperties}.}

\author[label1]{Oscar H. Ibarra\fnref{fn1}}
\address[label1]{Department of Computer Science\\ University of California, Santa Barbara, CA 93106, USA}
\ead[label1]{ibarra@cs.ucsb.edu}
\fntext[fn1]{Supported, in part, by
NSF Grant CCF-1117708 (Oscar H. Ibarra).}

\author[label2]{Ian McQuillan\fnref{fn2}}
\address[label2]{Department of Computer Science, University of Saskatchewan\\
Saskatoon, SK S7N 5C9, Canada}
\ead[label2]{mcquillan@cs.usask.ca}
\fntext[fn2]{Supported, in part, by Natural Sciences and Engineering Research Council of Canada Grant 2016-06172 (Ian McQuillan).}

\begin{abstract}
We look at nondeterministic finite automata augmented with multiple reversal-bounded counters
where, during an accepting computation, 
the behavior of the counters is specified by some fixed pattern. These patterns can serve as a useful ``bridge'' to other important automata and grammar models in the theoretical computer science literature, thereby helping in their study. Various pattern
behaviors are considered, together with characterizations and comparisons.
For example, one such pattern defines exactly
the smallest full trio containing all the bounded semilinear languages.
Another pattern defines the smallest full trio containing all the bounded
context-free languages.  The ``bridging'' to other families is then
applied, e.g.\ to certain Turing machine restrictions, as well as other
families. Certain general decidability properties are also
studied using this framework.
\end{abstract}


\end{frontmatter}

\section{Introduction}
\label{sec:intro}

Machines combining together a one-way read-only input augmented by one or more stores have been studied
considerably in the area of automata theory; for example, nondeterministic pushdown automata
($\NPDA$s) combine a one-way input plus a pushdown stack, which accept exactly the context-free languages
\cite{HU}. For such a machine model, it can be useful to restrict the ways in which the stores
can change. For example, an $\NPDA$ is {\em reversal-bounded} if there is a number $k$
where the number of changes between pushing and popping and vice versa on the pushdown is at most $k$. 
These machines characterize the ultralinear grammars \cite{GiSp68a} (when $k=1$, they characterize the linear grammars). Another example is with stack automata, which are more general than pushdown automata as they can also read from the inside of the stack in a two-way read-only fashion.
Two common restrictions of stack automata are non-erasing stack automata, and checking-stack automata
\cite{CheckingStack}; the first restricts stack automata so that no popping is allowed to occur, and the second  restricts them so that no popping is allowed to occur, plus after the machines start reading from the inside of the stack, they can no longer change its contents. All of these models can be viewed in terms of restricting the allowable sequences of operations that are allowed to occur on their stores. For stack automata, this was the approach taken in \cite{checkingstacksTCS}, which defines an {\em instruction language}, a language where each letter of the alphabet is associated with some operation on the store, and a machine is only allowed to follow computations where the store changes according to a word in the instruction language. Lastly, pushdown automata with only one pushdown symbol (plus a bottom-of-stack marker) act similarly to a counter. Although machines with two counters have the same power as Turing machines \cite{HU}, it is possible to restrict the operations of the stores to be reversal-bounded on each counter. This restriction defines the class $\NCM$ \cite{Ibarra1978}, which can be defined in terms of instruction languages as well \cite{StoreLanguages}.  The notion of instruction languages was built into arbitrary types of automata as well due to their usefulness in describing existing machine models in the literature \cite{StoreLanguages}. 

$\NCM$s in particular have been extensively studied in the literature theoretically \cite{Ibarra1978,Baker1974} as has their deterministic variant $\DCM$ \cite{DCMDeletion,DCMInsertion}. These machines have been applied to the areas of model-checking and verification \cite{HagueLin2011,IBARRA2002165}, timed automata \cite{Dang2001}, and Diophantine equations \cite{IbarraDang}. They also have desirable decidability properties, like polynomial time emptiness checking, where the number of counters is fixed \cite{Gurari1981220}.
Due to their utility, it is interesting to restrict them even further to see what families of languages can be accepted, and to examine their properties. This topic is pursued in this paper, where various families of instruction languages are examined together with the $\NCM$s that satisfy those instruction languages.

A language $L$ is bounded if $L \subseteq w_1^* \cdots w_k^*$, for non-empty words $w_1, \ldots, w_k$. Further,
$L$ is {\em bounded semilinear} if there exists a semilinear
set $Q \subseteq \mathbb{N}^k_0$ such that 
$L = \{w \mid w = w_1^{i_1} \cdots w_k^{i_k}, (i_1, \ldots, i_k) \in Q\}$ \cite{IbarraSeki}. 
It is known that every bounded
semilinear language can be accepted by a one-way
nondeterministic reversal-bounded multicounter machine
($\NCM$, \cite{Ibarra1978}). Also, every bounded language
accepted by an $\NCM$ can be accepted by a deterministic machine
in $\DCM$ \cite{IbarraSeki}. Thus,
every bounded semilinear language can be accepted by a $\DCM$.
Recently, in \cite{CIAA2016},
it was shown that in any language family $\LL$ 
that is a semilinear trio
(the Parikh image of each language is a semilinear set, and the family is closed
under $\lambda$-free homomorphism, inverse homomorphism, and 
intersection with regular languages), 
all bounded languages within $\LL$ are bounded semilinear, and can therefore be accepted by machines in $\NCM$ and, hence, in $\DCM$.
This implies that the equality problem, containment problem, and disjointness problem are decidable for bounded languages in $\LL$ (as long as the closure properties provide effective constructions) since they are decidable for $\DCM$ \cite{Ibarra1978}.
Hence, using instruction languages to restrict the operation of $\NCM$s has the potential to be useful in the study of arbitrary semilinear trios. We find that this is exactly the case, as restricting $\NCM$s with 
different instruction language families can create a ``bridge'' to other models of automata in the literature.

Informally, a $k$-counter machine $M$ 
is said to satisfy instruction
language $I \subseteq \{C_1,D_1, \ldots, C_k, D_k\}^*$ if, for every accepting computation of $M$, replacing
each increase of counter $i$ with $C_i$, and decrease of counter $i$ with $D_i$, gives a sequence in $I$.
Then, for a family of instruction languages ${\cal I}$, 
$\NCM({\cal I})$ is the family of $\NCM$ machines satisfying some $I \in {\cal I}$. 
Several interesting
instruction language families are defined and studied. 
For example, if one considers $\BD_i\LB_d$,
the family of instruction languages consisting of bounded increasing instructions
followed by letter-bounded decreasing instructions, then 
the family of languages accepted by machines in $\NCM(\BD_i\LB_d)$ is the smallest full trio containing all bounded  semilinear languages (and therefore, the smallest full trio containing all bounded languages from any semilinear trio).  
A characterization is also given of the smallest full trio containing the
bounded context-free languages by using
a natural subfamily of counter languages. Several other families are also defined and compared (results summarized in Figure
\ref{comparison}). 
For each, characterizations are given with a single language for each number of counters 
such that the families are the smallest full trios containing the languages. 
As a result of these characterizations,  a simple criterion is given for testing if the bounded languages within a given semilinear full trio
coincides with the bounded semilinear languages.

The usefulness of these characterizations using instructions languages is then demonstrated by considering several applications to interesting families, such as the multi-pushdown
languages \cite{multipushdown} and restricted types of Turing machines. The characterizations developed are used to show that the bounded languages
within each of these families are the same as those accepted by machines in $\DCM$. 
As the models themselves are quite different, they would be more difficult to compare
without the general techniques devised here.

\section{Preliminaries}
\label{sec:prelims}

In this paper, knowledge of automata and formal language theory are assumed; see 
\cite{HU} for an introduction.
Let $\Sigma$ be a finite alphabet. A word over $\Sigma$ is any
finite sequence of symbols from $\Sigma$, and $\Sigma^*$ 
(resp.\ $\Sigma^+$) is the set of all words (non-empty words)
over $\Sigma$, with the empty word denoted by $\lambda$. 
A language
is any $L \subseteq \Sigma^*$. 
The complement of $L$ with respect to $\Sigma^*$ is
$\overline{L} = \Sigma^* - L$. The shuffle of words $u,v\in \Sigma^*$ is
$u\shuffle v = \{u_1 v_1 \cdots u_n v_n \mid n\geq 1, u = u_1 \cdots u_n,
v = v_1 \cdots v_n, u_i, v_i \in \Sigma^*, 1 \leq i \leq n\}$, 
extended to languages $L_1 \shuffle L_2 = \{u \shuffle v \mid u \in L_1, v \in L_2\}$.
A language $L\subseteq \Sigma^*$ is bounded if there exist
$w_1, \ldots, w_k \in \Sigma^+$ such that $L \subseteq w_1^* \cdots w_k^*$,
and it is letter-bounded if $w_1, \ldots, w_k$ are letters. Furthermore,
$L$ is distinct-letter-bounded if each letter is distinct.

Let $\mathbb{N}$ be the set of positive integers and $\mathbb{N}_0 = \mathbb{N} \cup \{0\}$. 
A {\em linear set} is a set
$Q \subseteq \mathbb{N}_0^m$ if there exists $\vec{v_0},
\vec{v_1}, \ldots, \vec{v_n} \in \mathbb{N}_0^m$ such that
$Q = \{\vec{v_0} + i_1 \vec{v_1} + \cdots + i_n \vec{v_n} \mid
i_1, \ldots, i_n \in \mathbb{N}_0\}$. The vector
$\vec{v_0}$ is called the {\em constant}, and 
$\vec{v_1}, \ldots, \vec{v_n}$ are the {\em periods}.
A {\em semilinear set} is a finite union of linear sets.
Given an alphabet $\Sigma = \{a_1, \ldots, a_m\}$,
the length of a word $w\in \Sigma^*$ is denoted by $|w|$.
And, given $a \in \Sigma$, $|w|_a$ is the number of $a$'s in $w$.
The {\em Parikh image} of $w$ is
$\psi(w) = ( |w|_{a_1}, \ldots, |w|_{a_m})$, and the
Parikh image of a language $L$ is $\psi(L) = \{\psi(w) \mid w \in L\}$.
Also, $\alp(w) = \{ a \in \Sigma \mid |w|_a>0\}$.
A language $L$ is {\em semilinear} if $\psi(L)$ is a semilinear set.

As described in the Introduction, a language $L$ is {\em bounded semilinear} (called bounded Ginsburg semilinear in \cite{CIAA2016}) if there exists a semilinear
set $Q \subseteq \mathbb{N}^k_0$ such that 
$L = \{w \mid w = w_1^{i_1} \cdots w_k^{i_k}, (i_1, \ldots, i_k) \in Q\}$ \cite{IbarraSeki}. 
To contrast, a language $L \subseteq w_1^* \cdots w_k^*$ over alphabet $\Sigma$
is {\em bounded Parikh semilinear} if $L =\{ w \mid w = w_1^{i_1} \cdots w_k^{i_k},\mbox{~the Parikh map of $w$ is in~} Q\}$, where $Q$
is a semilinear set with $|\Sigma|$ components. In \cite{CIAA2016}, it was shown that 
the family
of bounded Parikh semilinear languages is a strict subset of the family of bounded semilinear languages (which
is a strict subset of the family of languages that are both bounded and semilinear).

Nondeterministic (deterministic resp.) finite automata are denoted by $\NFA$ ($\DFA$ resp.), both of which accept
the regular languages \cite{HU}.
For a class of machines ${\cal M}$, let $\LL({\cal M})$ be
the family of languages accepted by machines in ${\cal M}$.
Let $\LL(\CFL)$ be the family of context-free languages.
A {\em trio} (resp.\ {\em full trio}) is any family of languages closed under
$\lambda$-free homomorphism (resp.\ homomorphism), inverse homomorphism, and intersection
with regular languages. A {\em full semi-AFL} is a full trio closed
under union. Many well-known families of languages are trios, such as every family of the Chomsky hierarchy. Many important families are full trios  and full semi-AFLs as well, such as the families of regular and context-free languages. Given a language family $\LL$, $\LL\bd$ are the bounded languages in $\LL$.
Also, given $\LL$, the smallest full trio containing $\LL$ is denoted by $\fulltrio(\LL)$, and the smallest
full semi-AFL containing $\LL$ is denoted by $\fullsemiafl(\LL)$. The study of arbitrary families closed
under various language operations such as trio, semi-AFL and the more general AFL (which stands for
abstract family of languages) began by Ginsburg and Greibach \cite{GGr1}. A comprehensive
set of results appears in \cite{G75}.

A {\em one-way} $k$-counter machine \cite{Ibarra1978} is a tuple
$M = (k,Q,\Sigma,\lhd,\delta,q_0,F)$, where
$Q,\Sigma,\lhd,q_0,F$ are respectively the
finite set of states, input alphabet, right input end-marker
(which is unnecessary for nondeterministic machines and will largely
not be used in this paper), initial state, and final states.
The transition function is a partial function from $Q \times (\Sigma \cup \{\lhd,\lambda\}) \times \{0,1\}^k$ to the powerset of $Q \times \{-1,0,+1\}^k$, such that
$(p,d_1, \ldots, d_k) \in \delta(q,x, c_1, \ldots, c_k)$ and
$c_i = 0$ implies $d_i \geq 0$ to prevent negative values in the counters. Also, $M$ is deterministic if
$|\delta(q,a,c_1, \ldots, c_k) \cup \delta(q,\lambda,c_1, \ldots, c_k)| \leq 1$, for all $q \in Q, a \in \Sigma\cup \{\lhd\}, (c_1, \ldots, c_k) \in \{0,1\}^k$. A configuration of $M$ is a tuple $(q,w,i_1, \ldots , i_k)$
where $q$ is the current state, $w \in \Sigma^*\lhd \cup \{\lambda\}$ is
the remaining input, and $i_1, \ldots, i_k$ are the contents of the
counters. The derivation relation $\vdash_M$ is defined between
configurations, where 
$(q,xw, i_1, \ldots, i_k) \vdash_M (p,w, i_1+d_1, \ldots , i_k + d_k)$
if there is a transition $(p,d_1, \ldots, d_k) \in 
\delta(q,x, c_1, \ldots, c_k)$, where $c_j$ is $1$ if $i_j>0$, and 
$c_j = 0$ otherwise, if $i_j = 0$. 
Hence, a transition with $c_j = 0$ is applied if counter $j$ is empty, and a transition with $c_j =1$ is applied if counter $j$ is non-empty.
Let $\vdash_M^*$ be the 
reflexive, transitive closure of $\vdash_M$. $M$ accepts a word $w \in \Sigma^*$ if $(q_0, w\lhd, 0, \ldots, 0) \vdash_M^* (q_f, \lambda, i_1, \ldots, i_k), q_f \in F, i_1, \ldots, i_k \in \mathbb{N}_0$, and
the language of all words accepted by $M$ is denoted by $L(M)$.

Further, $M$ is $l$-reversal-bounded if,
in every accepting computation, the counter alternates between non-decreasing and non-increasing at most $l$ times. Often, labels from an alphabet $T$ will be associated
 to the transitions of $M$ bijectively, and then
$\vdash_M^t$ is written to represent the changing of configurations via transition $t$. This is generalized to derivations over words in $T^*$.

The class of one-way $l$-reversal-bounded $k$-counter
machines is denoted by $\NCM(k,l)$, and $\NCM$ is all reversal-bounded multicounter machines,
and replacing N with D gives the deterministic variant.


\

\section{Instruction Languages and Counter Machines}
\label{sec:instruction}

All bounded
languages in every semilinear trio are in $\LL(\NCM)$ \cite{CIAA2016}. Therefore, it is useful to consider 
subclasses of $\LL(\NCM)$ in order to determine more restricted
methods of computation where this property also holds.
Characterizations of the restricted families
are also possible, and these lead to simple methods
to determine the set of bounded languages within any semilinear full trio of interest.

First, restrictions of $\NCM$s are defined depending on the 
sequences of counter instructions that are allowed to occur.
For simplicity, these restrictions will only be defined on $\NCM$s that we will
call {\em well-formed}. A $k$-counter $\NCM$ $M$ is well-formed if
$M \in \NCM(k,1)$ (each counter does not increase again after decreasing) whereby all transitions change at most
one counter value per transition, and all counters decrease to zero
before accepting. Indeed, an $\NCM$ (or $\DCM$) can be assumed
without loss of generality
to be $1$-reversal-bounded by increasing the number of counters \cite{Ibarra1978}. 
Further, it is also
clear that all counters can be forced to change one counter value
at a time, and decrease to zero before accepting without loss
of generality. Thus, every language in $\LL(\NCM)$ can be
accepted by a well-formed $\NCM$.
Also, since only nondeterministic machines will be considered,
the input end-marker $\lhd$ will not be included.
 
Let $\Delta= \{C_1, D_1, C_2, D_2, \ldots \}$ be an infinite set of new symbols. In addition, for $k \geq 1$, let $\Delta_k = \{C_1,D_1, \ldots, C_k, D_k\}$, $\Delta_{(k,c)} = \{C_1, \ldots, C_k\}, \Delta_{(k,d)} = \{D_1, \ldots, D_k\}$.

Given a well-formed $k$-counter $\NCM$ machine $M = (k, Q, \Sigma, \lhd, \delta, q_0,F)$, let $T$ be a set of labels in bijective correspondence
with transitions of $M$. 
Define a homomorphism $h_{\Delta}$ from $T^*$ to $\Delta_k^*$
that maps every transition label associated with a transition that increases counter $i$ to $C_i$, maps every label
associated with a transition that decreases counter $i$ to $D_i$, and maps all labels associated with transitions that do not change
any counter to $\lambda$.
Also, define a homomorphism 
$h_{\Sigma}$ from $T^*$ to $\Sigma^*$ that maps every transition label $t$ associated with a transition that reads a letter
$a \in \Sigma$ to $a$, and erases all others.
An {\em instruction language} is any $I \subseteq \Delta_k^*$, for some $k \geq 1$.
It is said that $M$ {\em satisfies instruction language} $I \subseteq \Delta_k^*$
if every sequence of transitions $\alpha \in T^*$ corresponding to an
accepting computation --- that is $(q_0, w, 0, \ldots, 0)\vdash_M^{\alpha} (q, \lambda, 0, \ldots, 0), q$ a final state --- has
$h_{\Delta}(\alpha) \in I$. This means that $M$ satisfies instruction language $I$ if $I$ describes all possible sequences of counter increase and decrease instructions that can be performed in an accepting computation by $M$, with $C_i$ occurring for every increase of counter $i$ by one, and $D_i$ occurring for every decrease of counter $i$ by one.
Note that $I$ can contain additional sequences that do not get used.

Given a family of instruction languages ${\cal I}$ 
(therefore each $I \in {\cal I}$ is over $\Delta_k$, for some
$k \geq 1$), let $\NCM(k,{\cal I})$ be the subset of well-formed $k$-counter $\NCM$ machines that satisfy $I$ for some $I \in {\cal I}$ with $I \subseteq \Delta_k^*$; these are called the $k$-counter ${\cal I}$-instruction machines. The family of languages they accept,
$\LL(\NCM(k,{\cal I}))$, are called
the $k$-counter ${\cal I}$-instruction languages. Furthermore, $\NCM({\cal I}) = \bigcup_{k \geq 1} \NCM(k, {\cal I})$ are the ${\cal I}$-instruction machines, and $\LL(\NCM({\cal I})) = \bigcup_{k \geq 1} \LL(\NCM(k,{\cal I}))$ are the ${\cal I}$-instruction languages.
Instruction languages $I$ will only be considered
where, for all $w \in I$, every occurrence
of $C_i$ occurs before any occurrence of $D_i$, for all $i$, $1 \leq i \leq k$, which
is enough since every well-formed machine is $1$-reversal-bounded.

Some general properties of these restrictions will be studied
before examining some specific types.

\begin{proposition}
Given any family of instruction languages ${\cal I}$ over $\Delta_k$, $\LL(\NCM(k, {\cal I}))$ is a full trio. Furthermore, given any family of instruction languages ${\cal I}$, $\LL(\NCM({\cal I}))$ is a full trio.
\label{semiAFL}\end{proposition}
 
\begin{proof}
The standard proofs for closure under homomorphism and inverse homomorphism apply. The proof for intersection with regular languages
also works, as restricting the words of the language accepted can restrict the possible sequences of instructions
appearing in accepting computations, and the resulting sequences of instructions that can appear in an accepting computation 
will therefore be a subset of the instruction language of the original machine.
\qed
\end{proof}

Another definition is required. Given a language $I$ over $\Delta_k$, let $$I_{eq} = \{ w \mid w \in I, |w|_{C_i} = |w|_{D_i}, \mbox{every occurrence of $C_i$ occurs before any $D_i$, for~} 1 \leq i \leq k\}.$$
Further, given a family of instruction languages ${\cal I}$ over $\Delta$ 
then ${\cal I}_{eq}$ is the family of all languages
$I_{eq}$, where $I \in {\cal I}$.

\begin{proposition}
\label{generalSmallest}
Let ${\cal I}$ be a family of instruction languages where ${\cal I}$ is a subfamily
of the regular languages. Then $\LL(\NCM({\cal I})) = \fulltrio({\cal I}_{eq})$.
\end{proposition}
 
\begin{proof}
To see that ${\cal I}_{eq} \subseteq \LL(\NCM({\cal I}))$, let $I \in {\cal I}$ and let
$M$ be a $\DFA$ accepting $I \subseteq \Delta_k^*$. Create 
a well-formed $k$-counter machine $M'$ that accepts $I_{eq}$ as follows:
$M'$ simulates $M$ while adding to counter $i$ for every $C_i$ read,
and subtracting from counter $i$ for every $D_i$ read (never adding after
subtracting), accepting if $M$ does, and if all counters end at zero. Thus, $M'$ accepts all
words of $I$ with an equal number of $C_i$'s as $D_i$'s, for each $i$
where all $C_i$'s occur before any $D_i$.
This is exactly $I_{eq}$. Also, $M'$ satisfies $I_{eq}$ and $I$.
Hence ${\cal I}_{eq} \subseteq \LL(\NCM({\cal I}))$. It follows from Proposition \ref{semiAFL} that $\LL(\NCM({\cal I}))$ is a full trio. Thus, $\fulltrio({\cal I}_{eq}) \subseteq \LL(\NCM({\cal I}))$

Next it will be verified that $\LL(\NCM({\cal I})) \subseteq \fulltrio({\cal I}_{eq})$.
Consider a machine $M = (k,Q,\Sigma,\lhd,\delta,q_0,F) \in \NCM({\cal I})$ with $k$ counters that satisfies instruction language $I \in {\cal I}$. 

Let $\Gamma \subseteq Q \times (\Sigma \cup \{\lambda\}) \times
(\Delta_k \cup \{0\}) \times Q$, where $(q,a,X,p) \in \Gamma$ if and only if there is a transition $t$ of
$\delta$ from $q$ to $p$ on $a \in \Sigma \cup \{\lambda\}$ with 
$$X = 
\begin{cases} C_i & \mbox{if~} t \mbox{~increases counter~} i,\\
D_i & \mbox{if~} t \mbox{~decreases counter~} i,\\
0 & \mbox{if~} t \mbox{~does not change any counter.}
\end{cases}
$$
Let
$g$ be a homomorphism from $\Gamma^*$ to $\Delta_k^*$
that maps
$(q,a,X,p)$ to $X$, if $X \in \Delta_k$, and to $\lambda$ if $X=0$. 
Both of these types of symbols can be created from
transitions defined on any counter value ($0$ or positive). 
It is said that symbol
$(q,a,X,p), X \in \Delta_k \cup \{0\}$ 
is defined on counter $i$ positive if it was created above from a transition defined on counter $i$ being positive. It is said that
the symbol is defined on counter $i$ being zero if it was created
from a transition on counter $i$ being zero. Such a symbol could
be defined on counter $i$ being both $0$ and positive if more than one transition creates one symbol.

Create regular language $R \subseteq \Gamma^*$ consisting of all words $y_0 y_1 \cdots y_n$ such that 
$y_i = (p_i, a_{i+1}, X_{i+1}, p_{i+1})$, $p_0 = q_0, p_{n+1} \in F,$
for each $i$, $1 \leq i \leq k$, if $j$ is the smallest $r$ in $\{0, \ldots, n\}$ such that $X_r = C_i$ and if $l$ is the largest $s$ in $\{0, \ldots, n\}$ such that $X_s = D_i$, then $y_0, \ldots, y_j$ are defined on counter $i$ zero, $y_{j+1}, \ldots, y_l$
are defined on counter $i$ positive, and 
$y_{l+1}, \ldots, y_n$ are defined on counter $i$ zero.

Let $h$ be a homomorphism from $\Gamma^*$ to $\Sigma^*$
such that $h$ projects onto the second
component. It is clear that $L(M) = h(g^{-1}(I_{eq})\cap R)$
since $g^{-1}(I_{eq})\cap R$ consists of all words of $R$ with an
equal number of $C_i$'s as $D_i$'s, for each $i$, $1 \leq i \leq k$.

Hence, $\LL(\NCM({\cal I})) = \fulltrio({\cal I}_{eq})$.
\qed
\end{proof}

Next, several specific instruction language families will be defined that can be used to define
interesting subfamilies of $\LL(\NCM)$. 
Each instruction language family is a  natural subfamily of the regular languages (therefore satisfying
Proposition \ref{generalSmallest}).

\begin{definition}
\label{def:families}
Consider the instruction language families:
\begin{itemize}

\item $\LB_i\LB_d =  \{ I = YZ \mid k \geq 1, Y = a_1^* \cdots a_m^*, a_i \in \Delta_{(k,c)}, 1 \leq i \leq m, Z= b_1^* \cdots b_n^*, b_j \in \Delta_{(k,d)}, 1 \leq j \leq n\}$, (letter-bounded-increasing/letter-bounded-decreasing instructions),


\item $\StLB = \{ I \mid k \geq 1, I = a_1^* \cdots a_m^*, a_i \in \Delta_k,  1 \leq i \leq m, \mbox{~there is no~} 1 \leq l < l'<j<j' \leq m \mbox{~such that~} a_l = C_r, a_{l'} =C_s, a_j = D_r,  a_{j'} = D_s, r \neq s \}$,  (stratified-letter-bounded
instructions),


\item $\LB = \{ I \mid k \geq 1, I = a_1^* \cdots a_m^*, a_i \in \Delta_k, 1 \leq i \leq m\}$, (letter-bounded instructions),


\item $\BD_i \LB_d = \{ I = YZ \mid k \geq 1, Y = w_1^* \cdots w_m^*, w_i \in \Delta_{(k,c)}^*, 1 \leq i \leq m, Z= a_1^* \cdots a_n^*, a_j \in \Delta_{(k,d)}, 1 \leq j \leq n\}$, (bounded-increasing/letter-bounded-decreasing instructions),


\item $\LB_i \BD_d = \{ I = YZ \mid k \geq 1, Y = a_1^* \cdots a_m^*, a_i \in \Delta_{(k,c)}, 1 \leq i \leq m, Z = w_1^* \cdots w_n^*, w_j \in \Delta_{(k,d)}^*, 1 \leq j \leq n\}$, (letter-bounded-increasing/bounded-decreasing instructions),


\item $\BD = \{ I \mid k \geq 1, I = w_1^* \cdots w_m^*, w_i \in \Delta_k^*, 1 \leq i \leq m\}$, (bounded instructions),


\item $\LB_d = \{ I \mid k \geq 1, I = Y \shuffle Z, Y = \Delta_{(k,c)}^*, Z = a_1^* \cdots a_n^*, a_j \in \Delta_{(k,d)}, 1 \leq j \leq n\}$, (letter-bounded-decreasing instructions),


\item $\LB_i = \{ I \mid k \geq 1, I = Y \shuffle Z, Y = a_1^* \cdots a_m^*, a_i \in \Delta_{(k,c)}, 1 \leq i \leq m, Z = \Delta_{(k,d)}^*, \}$, (letter-bounded
increasing instructions),


\item $\LB_{\cup} = \LB_d \cup \LB_i$, (either letter-bounded-decreasing
or letter-bounded-increasing instructions),


\item $\ALL = \{ I \mid k \geq 1, I = \Delta_k^*\}$.
\end{itemize}

\end{definition}

For example, for every well-formed $\NCM$ machine $M$ where the counters are increased and decreased according to some bounded instruction language, then there is an instruction language $I$ such that $M$ satisfies $I$, and $I \in \BD$, and $L(M) \in \LL(\NCM(\BD))$. Even though not all instructions in $I$ are necessarily used, the instructions used will be a subset of $I$ since the instructions used are a subset of a bounded language.
It is also clear that $\LL(\NCM) = \LL(\NCM(\ALL))$ since all sequences of instructions are permitted in $\ALL$. In the case of the latter five families where it is not explicitly stated, only languages $I$ in these families
will be considered where every occurrence of $C_i$ occurs before any $D_i$, for all $i$, as only well-formed machines are used.

\begin{example} \label{ex2}
Let $L =  \{ u a^i  v  b^j  w  a^i x  b^j y ~|~  i, j > 0,
u,v,w,x,y \in \{0,1\}^* \}$.
One can construct a well-formed 2-counter machine $M$ to
accept $L$ where, on input $u a^i  v  b^j  w  a^{i'} x  b^{j'} y$, $M$ increases counter 1 $i$ times, then increases counter 2 $j$ times, then
decreases counter 1 verifying that $i = i'$, then decreases counter 2 verifying that $j=j'$.
This machine satisfies instruction language
$C_1^* C_2^* D_1^* D_2^*$, which is a subset of some instruction
language in every family in Definition \ref{def:families} except
for $\StLB$, and therefore $L \in \LL(\NCM({\cal I}))$ for each of these families ${\cal I}$.
This does not immediately imply that $L \notin \LL(\NCM(\StLB))$ as there could possibly be another machine
accepting $L$ satisfying some instruction language in $\StLB$.
\end{example}

\begin{example} \label{ex3}
Let $L = \{a^{2+i+2j} b^{3+2i+5j} \mid i, j \ge 0 \}$.
The Parikh image of $L$
is a linear set $Q = \{(2,3) + (1,2)i + (2,5)j ~|~ i, j \ge 0\}$.
Indeed, 
$L$ can be accepted by a well-formed 4-counter $\NCM$ $M$ as follows,
when given input $a^m b^n$:
first, 
on $\lambda$-moves, $M$ increments counters 1 and 2 alternatingly
a nondeterministically guessed number of times $i \ge 0$, then
on $\lambda$-moves, increments counters 3 and 4 alternatingly
a nondeterministically guessed number of times $j \ge 0$.
From there, 
$M$ verifies that $m = 2 + i +2j$ by
reading 2 $a$'s, then decrementing 
counter 1 to zero while reading one $a$ for each decrease, and then decrease counter 3 to zero to check that the
remaining number of $a$'s is equal to 2 times the value of counter 3. Finally, $M$ checks and accepts if $n = 3+ 2i +5j$ by first
reading 3 $b$'s and decrementing counter 2 and then 4 appropriately.

The instructions of $M$ as constructed are a subset of
$I= (C_1C_2)^* (C_3C_4)^* D_1^* D_3^* D_2^*D_4^*$. This is a
subset of some language in
each of $\BD_i\LB_d, \BD, \LB_d$ but not the other families, 
and therefore $M$ is in each of $\NCM(\BD_i\LB_d), \NCM(\BD),
\NCM(\LB_d)$. Even though $M$ is not in the other classes of machines 
such as $\NCM(\LB_i)$, it is possible for $L(M)$ to be in $\LL(\NCM(\LB_i))$ (using some other machine that accepts the same language). Indeed, we
will see that $L(M)$ is also in $\LL(\NCM(\LB_i\BD_d))$ and $\LL(\NCM(\LB_i))$.
\end{example}

\begin{example} \label{ex4a}
Let $L_1 = \{w \#a^i b^j \mid |w|_a = i, |w|_b = j\}$.

Construct $M_1$ that on input $w\# a^i b^j$, reads $w$ while adding to the first counter for every $a$
read, and adding to the second counter for every $b$ read. After $\#$,
$M_1$ subtracts from counter $1$ for every $a$ read, then when
reaching $b$, it switches to decreasing the second counter
for every $b$ read. Therefore,
it satisfies language $\{C_1, C_2\}^* D_1^* D_2^*$
which is indeed a subset of $(\{C_1, C_2\}^* \shuffle D_1^* D_2^*)
\in \LB_d$.

Now let $L_2 = L_1^R$.  We conjecture that $L_2$  is not in $\LL(\NCM(\LB_d))$,
but we can construct a machine $M_2 \in \NCM(\LB_i)$ to accept $L_2$.
($M_2$, when given $b^j a^i \# w$, stores $i$ and $j$ in
two counters and then checks by decrementing the counters that
$|w|_a = i$ and $|w|_b = j$.)
Similarly, we conjecture that $L_1$ is not in $\LL(\NCM(\LB_i))$.
Obviously, $L_1$ and $L_2$ are both in $\LL(\NCM(\LB_{\cup}))$.

\end{example}

\begin{example} \label{ex4}
Let $L = \{ w  \mid  w \in \{a,b\}^+, |w|_a = |w|_b > 0\}$.
$L$ can be accepted by an $\NCM$ which uses two counters
that increments counter 1 (resp. counter 2) whenever
it sees an $a$ (resp., $b$). Then it decrements counter 1 and counter 2 simultaneously and
accepts if they reach zero at the same time.
This counter usage does not
have a pattern in any of the restrictions above except $\ALL$. It is quite unlikely that $L(M) \in \LL(\NCM({\cal I}))$ for any of the families in the definition above except the full $\LL(\NCM(\ALL)) = \LL(\NCM)$.
\end{example}

Every family ${\cal I}$ in Definition \ref{def:families} is
a subfamily of the regular languages. Therefore, by Proposition \ref{generalSmallest}, the following can be shown
by proving closure under union:
\begin{proposition}
\label{eachSmallest}
Let ${\cal I}$ be any family of instruction languages from
Definition \ref{def:families}. Then
$\LL(\NCM({\cal I})) = \fullsemiafl({\cal I}_{eq})$.
\end{proposition}
 
\begin{proof}
It suffices to show closure under union for each family.
For $\LB_i\LB_d$, given two machines $M_1, M_2$ with $k_1, k_2$
counters respectively, a $k_1 + k_2$ counter machine $M$ can be built
where $M$ adds to counters according to $M_1$ using the first
$k_1$ counters, then decreasing according to $M_1$,  or the same with $M_2$ on the remaining
counters. It is clear that this
gives an instruction language that is a subset of the increasing
pattern of $M_1$ followed by the increasing pattern of $M_2$, 
followed by the decreasing pattern of $M_2$, then $M_1$. This is 
letter-bounded insertion followed by letter-bounded deletion
behavior. The same construction works in all other cases.
\qed
\end{proof}

As a corollary, when considering the instruction languages of $\ALL$
(thus, the instructions are totally arbitrary), and for $i \geq 1$, let $L_i = \{C_i^n D_i^n \mid n \geq 0\}$, then
$\ALL_{eq} = \{ I \mid I = L_1 \shuffle L_2 \shuffle \cdots \shuffle L_k,  k \geq 1\}$.
Hence, $\LL(\NCM) = \fulltrio(\ALL_{eq})$, by Proposition \ref{generalSmallest}.
Or, it could be stated as follows:
\begin{corollary} \cite{Baker1974,G78}
$\LL(\NCM)$ is the smallest shuffle or intersection closed full trio containing $\{a^n b^n \mid n \geq 0\}$.
\label{NCMintersection}
\end{corollary}
This is already known for intersection \cite{Baker1974,G78}, and it is known that for trios, a family is closed under intersection if and only if it is closed under shuffle (Exercise 5.5.6 of \cite{G75}).

Since $\{a^n b^n \mid n \geq 0 \}$ is in $\LL(\NCM({\cal I}))$ for
all ${\cal I}$ in Definition \ref{def:families}, the following is also immediate from Corollary \ref{NCMintersection}:
\begin{corollary}
\label{smallestIntersection}
For all ${\cal I}$ in Definition \ref{def:families}, $\LL(\NCM)$ is the smallest shuffle or intersection closed full trio containing $\LL(\NCM({\cal I}))$.
\end{corollary}
Thus, any instruction family ${\cal I}$ whereby $\LL(\NCM({\cal I})) \subsetneq \LL(\NCM)$ and $\{a^n b^n \mid n \geq 0\} \in \LL(\NCM({\cal I}))$ is immediately not closed under
intersection and shuffle.

The next lemma will be proven regarding many of the instruction language families showing that letter-bounded instructions can
be assumed to be distinct-letter-bounded, and for bounded languages, for each letter in $\Delta_k$ in the words to only appear once.

First, a definition is needed. For each of the instruction families of Definition \ref{def:families}, place
an underline below each $\LB$ if the letter-bounded language is forced to have each letter occur exactly once (and therefore be distinct-letter-bounded), and place an underline below $\BD$ if
each letter $a \in \Delta_k$ appears exactly once within the words $w_1, \ldots, w_m$. Thus, as an example, $\underline{\LB}_i \underline{\BD}_d$
is the subset of $\LB_i\BD_d$ equal to 
$\{ I = YZ \mid k \geq 1, Y = a_1^* \cdots a_k^*, a_i \in \Delta_{(k,c)}, |a_1 \cdots a_k|_a = 1, \mbox{~for all~} a \in \Delta_{(k,c)}, Z = w_1^* \cdots w_n^*, w_i \in \Delta_{(k,d)}^*, 1 \leq j \leq n, |w_1 w_2 \cdots w_n|_a = 1, \mbox{~for all~} a \in \Delta_{(k,d)}\}$. Thus, each letter appears
exactly once in the words or letters.
The construction uses multiple new instruction letters and counters, in order to allow each letter to only appear once.
This provides an interesting normal form for many types of counter machines. It is also quite useful and is a key towards separating the subfamilies of $\LL(\NCM)$.

\begin{lemma} \label{distinct}
The following are true:
\begin{tabbing}
$\LL(\NCM(\LB_i\LB_d)) = \LL(\NCM(\underline{\LB}_i\underline{\LB}_d))$, \ \=
$\LL(\NCM(\LB)) = \LL(\NCM(\underline{\LB}))$, \ \ \ \= $\LL(\NCM(\StLB)) = \LL(\NCM(\underline{\StLB}))$.\\
$\LL(\NCM(\LB_i\BD_d)) = \LL(\NCM(\underline{\LB}_i\underline{\BD}_d))$, \>
$\LL(\NCM(\LB_{d})) = \LL(\NCM(\underline{\LB}_{d}))$, \\
$\LL(\NCM(\BD_i\LB_d)) = \LL(\NCM(\underline{\BD}_i\underline{\LB}_d))$, \>
$\LL(\NCM(\LB_i)) = \LL(\NCM(\underline{\LB}_i))$.
\end{tabbing}

\label{LBdistinct}
\end{lemma}
 
\begin{proof}
First, consider the case for $\LB$. Let $\LB_l$ contain all languages $I = a_1^* \cdots a_n^*$, $a_1, \ldots, a_n \in \Delta_k$ for some $k$ where at most $l$ elements
$x \in \{1, \ldots, k\}$ have either $C_i$ or $D_i$ occur more than one time in $a_1, \ldots, a_n$. Thus, $\LB_0 = \underline\LB$.
The proof will proceed inductively by taking any $I = a_1^* \cdots a_n^* \in \LB_l, l>1$ where 
$a_i \in \Delta_k$, $1 \leq i \leq k$, and showing $I_{eq} \in \LL(\NCM(\LB_{l-1}))$. 
Thus, $(\LB_l)_{eq} \subseteq \LL(\NCM(\LB_{l-1}))$,
and $\LL(\NCM(\LB_{l})) = \fulltrio((\LB_l)_{eq})\subseteq \LL(\NCM(\LB_{l-1}))$
by Proposition \ref{generalSmallest}.
The result would then follow. 

Let $I = a_1^* \cdots a_n^* \in \LB_l, l>1, a_i \in \Delta_k, 1 \leq i \leq n$, where either $C_x$ or $D_x$ occurs
multiple times in $a_1, \ldots, a_n$, for some $x \in \{1, \ldots, k\}$. A machine 
$M \in \NCM(\LB_{l-1})$ accepting $I_{eq}$ is built as follows:
For each $y \in \{1, \ldots, k\}$, let 
\begin{equation} f(y) = i_{1,y}, \ldots, i_{\alpha_y,y},
\label{xup}
\end{equation}
be the sequence of 
all positions where $a_{i_p,y} = C_y$,
$1 \leq p \leq \alpha_y$ and $1 \leq i_{1,y} < \cdots < i_{\alpha_y,y} \leq n$, and let 
\begin{equation}
g(y) = j_{1,y}, \ldots, j_{\beta_y,y},
\label{xdown}
\end{equation} 
be all positions
where $a_{j_q,y} = D_y$, $1 \leq q \leq \beta_y$ and $1 \leq j_{1,y}< \cdots < j_{\beta_y,y} \leq n$. 
Given $w = a_1^{\gamma_1} \cdots a_n^{\gamma_n} \in I$, then $M$ must accept $w$ if and only if 
\begin{equation}
\label{sumup}
\mbox{for each~} y \in \{1, \ldots, k\}, \gamma_{i_1,y} + \cdots + \gamma_{i_{\alpha_y,y}} = \gamma_{j_{1,y}} + \cdots + \gamma_{j_{\beta_y,y}}.
\end{equation} 
We will construct $M$ to count for all $C_y$'s and $D_y$'s where $y \neq x$ in the obvious way (using $k-1$ counters). However, $M$ will
count $C_x$'s and $D_x$'s in a different way. Let $f(x) = i_1, \ldots, i_{\alpha}$ and $g(x) = j_1, \ldots, j_{\beta}$.


In $M$, it uses $\alpha \cdot \beta$ new counters to count the $C_x$'s and $D_x$'s, which are referred to by 
$(p, q), 1 \leq p \leq \alpha, 1 \leq q \leq \beta$,
and the characters $C_{(p,q)}$ (resp.\ $D_{(p,q)}$) will be used to represent the instruction character while increasing (resp.\ decreasing)
counter $(p,q)$; these can easily be replaced with consecutive
numbered characters in $\Delta_m$ at the end of the procedure. 
In section $i_p$, for $1 \leq p \leq \alpha$, when reading $a_{i_p}^{\gamma_{i_p}}$ (with $a_{i_p} = C_x$), $M$ uses and increases counter
$(p,1)$ for every $a_{i_p}$ read until some nondeterministically chosen spot where $M$
switches to and increases from counter $(p, 2)$ (while reading the same section $i_p$), etc., through counter $(p, \beta)$.
Then, when reading $a_{j_q}^{\gamma_{j_q}}$ (with $a_{j_q} = D_x$) for $1 \leq q \leq \beta$, $M$ decreases from counters $(1,q)$ until empty, then $(2,q)$, etc.\
until counter $(\alpha, q)$ is empty.

The sequence of instructions of $M$ in 
every accepting computation 
is therefore in $d_1^* \cdots d_l^*$, where the sequence $d_1, \ldots, d_l$ is obtained from $a_1, \ldots, a_n$ by replacing each occurrence of $C_x$ at position $i_p, 1 \leq p \leq \alpha$ by 
$C_{(p,1)}, \ldots, C_{(p,\beta)}$, and replacing each occurrence
of $D_x$ at position $j_q, 1 \leq q \leq \beta$ by
$D_{(1,q)}, \ldots, D_{(\alpha,q)}$.

Let $w = a_1^{\gamma_1} \cdots a_n^{\gamma_n} \in I_{eq}$. 
Indeed, $\gamma_{i_1} + \cdots + \gamma_{i_{\alpha}} = \gamma_{j_1} + \cdots + \gamma_{j_{\beta}}$.
Then $w$ can be accepted in $M$ as follows: for each section 
$i_p$, for $1 \leq p \leq \alpha$,
$M$ reads $a_{i_p} = C_x$'s while it adds to counters $(p,1), \ldots, (p,\beta)$ by amounts
$\gamma_{(p,1)}, \ldots, \gamma_{(p,\beta)}$ respectively (these
amounts determined in the algorithm below),
and 
for each section $j_q$, for $1 \leq j \leq \beta$,
$M$ reads $a_{j_q} = D_x$'s while it subtracts from counters $(1,q), \ldots, (\alpha, q)$
by amounts $\gamma_{(1,q)}, \ldots, \gamma_{(\alpha,q)}$ respectively,
such that, for each $1 \leq p \leq \alpha, 1 \leq q \leq \beta$,
the following will be true:
\begin{eqnarray}
\gamma_{i_p}&=& \gamma_{(p,1)} +  \cdots + \gamma_{(p,\beta)} , \label{pequation} \\
\gamma_{j_q}&=& \gamma_{(1,q)} + \cdots + \gamma_{(\alpha,q)} , \label{qequation} 
\end{eqnarray}
If Equations (\ref{pequation}) and (\ref{qequation}) are true (they will be verified to be true when the amounts are chosen via the algorithm below), then $M$ increases and decreases the same amount
and can therefore accept, but
where each new counter is increased and decreased in exactly one
section.

Intuitively, the situation can be visualized as follows:
$$\begin{array}{cccc|ccccc}
\gamma_{i_1}, & \ldots,  & \gamma_{i_{\alpha}} &&& \gamma_{j_1}, & \ldots, & \gamma_{j_{\beta}}\\\hline
\gamma_{(1,1)} & & \gamma_{(\alpha,1)} &&& \gamma_{(1,1)} & \cdots & \gamma_{(1,\beta)}\\
\vdots & & \vdots & && &\\
\gamma_{(1,\beta)} && \gamma_{(\alpha,\beta)} &&& \gamma_{(\alpha,1)} & \cdots & \gamma_{(\alpha,\beta)}
\end{array}$$
These amounts can be simulated in a ``greedy'' fashion, by
the following algorithm where $\gamma_{(p,q)}$ is the output,
for all $1 \leq p \leq \alpha, 1 \leq q \leq \beta$:
\begin{code}
\noindent
\t\t let $X(p) = \gamma_{i_p}, 1 \leq p \leq \alpha$; let $Y(q) = \gamma_{j_q}, 1 \leq q \leq \beta$;\\
\t\t let $p = 1; q=1; \gamma_{(p',q')} = 0, \forall \ p', q', 1 \leq p' \leq \alpha , 1 \leq q' \leq \beta$;\\
\t\t while ($p \leq \alpha$ and $q \leq \beta$) \linelabel{loop}\\
\t\t\t\t $\gamma_{(p,q)} = \min\{ X(p), Y(q) \}$; \linelabel{minline}\\
\t\t\t\t $X(p) = X(p) - \gamma_{(p,q)}$; \linelabel{setX}\\
\t\t\t\t $Y(q) = Y(q) - \gamma_{(p,q)}$; \linelabel{setY}\\
\t\t\t\t if ($X(p) = 0)$ then $p = p+1$; \linelabel{pinc}\\
\t\t\t\t if ($Y(q) = 0)$ then $q = q+1$;
\end{code}
In this algorithm, $X$ and $Y$ are initialized to hold
$\gamma_{i_p}$ and $\gamma_{j_q}$, for each $1 \leq p \leq \alpha,
1 \leq q \leq \beta$. And, as amounts from each are added
to various ``counters'' $\gamma_{(p,q)}$ in line \lineref{minline},
these same amounts are simultaneously
reduced from $X(p)$ and $Y(q)$ in lines 
\lineref{setX} and \lineref{setY} until they are zero.

It will be shown by induction that each time line \lineref{loop} is
executed, 
$X(p) = \gamma_{i_{p}} - \gamma_{(p,q-1)}- \cdots - \gamma_{(p,1)}$ and $Y(q) = \gamma_{j_{q}} - \gamma_{(p-1,q)} - \cdots - \gamma_{(1,q)}$,
and after line \lineref{setY} is executed,
$X(p) = \gamma_{i_{p}} - \gamma_{(p,q)}- \cdots - \gamma_{(p,1)}$ and $Y(q) = \gamma_{j_{q}} - \gamma_{(p,q)} - \cdots - \gamma_{(1,q)}$.

The base case, when $p = 1 = q$ the first time line \ref{loop} is
executed is true because 
$X(1) = \gamma_{i_1}$ and $Y(1) = \gamma_{j_1}$.

Assume it is true at some iteration when reaching
line \lineref{loop}, with $p \leq \alpha$
and $q \leq \beta$. Then
$\gamma_{(p,q)} = \min\{ \gamma_{i_p} - \gamma_{(p,q-1)} - \cdots - \gamma_{(p,1)}, \gamma_{j_q}- \gamma_{(p-1,q)} - \cdots - \gamma_{(1,q)} \}$ in line \lineref{minline}. Assume the first is minimal and the second is not. Then, in line \lineref{minline},
$X(p) = \gamma_{(p,q)} = \gamma_{i_p} - \gamma_{(p,q-1)} - \cdots - \gamma_{(p,1)}$, thus $X(p)$ is zero after line \lineref{setX},
and indeed $0 = \gamma_{i_p} - \gamma_{(p,q)} - \cdots - \gamma_{(p,1)}$,
which is what we want by induction. Also,
after line \lineref{setY}, $Y(q) = \gamma_{j_q}-\gamma_{(p,q)} - \cdots - \gamma_{(1,q)}$. Then, $p$ is increased in line \lineref{pinc},
and when reaching line \lineref{loop} again, $X(p)$ (where $p$ has been increased) is equal to $\gamma_{i_p}$
which is what we want since $\gamma_{(p,q-1)}, \ldots, \gamma_{(p,1)}$
are all zero.
Furthermore, $Y(q) = \gamma_{j_q}-\gamma_{(p-1,q)} - \cdots - \gamma_{(1,q)}$ since $q$ was not increased from line \lineref{setY}
of the previous iteration.
Similarly when the second case is (or both are) minimal.

Since $\gamma_{i_1} + \cdots + \gamma_{i_{\alpha}} = \gamma_{j_1} + \cdots + \gamma_{j_{\beta}}$, and the same
values are subtracted from some $X(p)$ and $Y(q)$ at each step, all
$X(p)$ and $Y(q)$ must decrease to $0$, and the final iteration has
to occur when $p = \alpha, q = \beta$, and in this case $X(p) = Y(q)$.
Further, for each $p,q$ in the loop where $X(p)$ is set to $0$,
$\gamma_{i_p}$ is the sum of $\gamma_{(p,q)}, \ldots, \gamma_{(p,1)}$,
and since $p$ is increased and never decreased again,
$\gamma_{(p,q+1)}, \ldots, \gamma_{(p,\beta)}$ are always $0$.
Thus, Equation (\ref{pequation}) is true. Similarly when $Y(q)$ hits zero, the same idea
demonstrates that Equation (\ref{qequation}) is true.
Thus, since $M$ calculates these values
$\gamma_{(p,q)}$ nondeterministically, they can be set to the
amounts calculated by this algorithm. Hence, $w \in L(M)$.

Let $w \in L(M)$. It is evident that $w \in I_{eq}$ since $M$ is reading $C_x$ for every
increase of counter $(p,q)$, and is reading $D_x$ for every decrease of counter $(p,q)$. Thus, $w \in I_{eq}$.

Thus, $I_{eq} = L(M)$.
The result follows by induction.

It is clear that the procedure works works identically for 
$\LB_i \LB_d$. For $\LB_d$, it is simpler, since there is no
restrictions on the increasing instructions, and similarly for
$\LB_i$.

Next, we will address the case for $\StLB$.
First, we show the following, which is helpful:
\begin{claim}
$\LL(\NCM(\underline{\StLB}))$ is closed under union.
\end{claim}
\begin{proof}
Take $M_1$ that satisfies instruction language $I_1 = a_1^* \cdots a_n^* \subseteq \Delta_k^*$, and $M_2$
that satisfies instruction language $I_2 = b_1^* \cdots b_m^* \subseteq \Delta_l^*$. 
Let $I_2'$ be obtained from $I_2$ by replacing $C_i$ ($D_i$) with $C_{i+k}$ ($D_{i+k}$). Consider $M$ satisfying instruction
language $I = I_1 I_2' \subseteq \Delta_{k+l}^*$ where $M$ either simulates $M_1$, or $M_2$ using
counter $i+k$ to simulate $i$. Then $I \in \underline{\StLB}$ if and only if $I_1$ and $I_2$ are in $\underline{\StLB}$, and $L(M) = L(M_1) \cup L(M_2)$.
\qed
\end{proof}

Let $I = \{a_1^{\gamma_1} \cdots a_n^{\gamma_n} \mid$ there is no $1 \leq l <l'< j <j' \leq n$ such that
$a_l = C_s, a_{l'} = C_r, a_j = D_s, a_{j'} = D_r\}  \in \StLB$.
For all $y \in \{1, \ldots, k\}$, let $f(y)$ and $g(y)$ be as in Equations (\ref{xup}) and (\ref{xdown}). We will build $M \in \NCM(\underline{\StLB})$ that accepts $I_{eq}$. 
It accepts if and only if Equation (\ref{sumup}) is true for all $w = a_1^{\gamma_1} \cdots a_n^{\gamma_n} \in I$. 
Let $x \in \{1, \ldots, k\}$ such that either $C_x$ or $D_x$ occurs multiple times in $a_1, \ldots, a_n$, and let $f(x) = i_1,\ldots, i_{\alpha}, g(x) = j_1, \ldots, j_{\beta}$.
For each $w \in I$ with $w = a_1^{\gamma_1} \cdots a_n^{\gamma_n}$,
either  $\gamma_{i_1} \leq \gamma_{j_{\beta}}$, or
$\gamma_{i_1} > \gamma_{j_{\beta}}$. Of course, one could partition $I_{eq}$ into two languages $I_1$ and $I_2$ accepted by two machines $M_1$ and $M_2$ where the first only contains
words where $\gamma_{i_1} \leq \gamma_{j_{\beta}}$ and the latter only contains those where $\gamma_{i_1} > \gamma_{j_{\beta}}$.
In the first case, the entirety of $a_{i_1}^{\gamma_{i_1}}$ can be read by some new counter $C_{(1,x)}$ which then can be emptied when reading the 
last $\gamma_{i_1} \leq \gamma_{j_{\beta}}$ characters of $a_{j_{\beta}}^{\gamma_{j_{\beta}}}$. This leaves 
$\gamma_{i_2}, \ldots, \gamma_{i_{\alpha}}, \gamma_{j_1}, \ldots, \gamma_{j_{\beta-1}}, \gamma_{j_{\beta}}'$,
where $\gamma_{j_{\beta}} = \gamma_{i_1} + \gamma_{j_{\beta}}'$ in terms of $C_x$'s and $D_x$'s. 
(The latter case where $\gamma_{i_1}>\gamma_{j_{\beta}}$ could be accepted by $M_2$ that has a counter
$C_{(1,x)}$ that reads $a_{i_1}^{\gamma_{j_{\beta}}}$ while increasing $C_{(1,x)}$ and $C_{(1,x)}$ empties entirely when reaching the last section; this leaves
$\gamma_{i_1}',\gamma_{i_2}, \ldots, \gamma_{i_{\alpha}},\gamma_{j_1},\ldots, \gamma_{j_{\beta-1}}$ where $\gamma_{i_1} = \gamma_{i_1}' + \gamma_{j_{\beta}}$ in terms of $C_x$'s and $D_x$'s).
Then machine $M_1$ can be built accepting $I_1$ where $M_1$ satisfies the instruction languages where the first $C_x$ in $I$ is replaced with $C_{(1,x)}$ and the 
last $D_x$ is replaced by $D_x D_{(1,x)}$, and $M_2$ accepting $I_2$ satisfies the instruction language where the first
$C_x$ is replaced by $C_{(1,x)} C_x$, and the last $D_x$ is replaced by $D_{(1,x)}$.
From both cases, this removes one section, and we can repeat this procedure inductively with the remaining sections until
finished. Notice that after doing this, in each case, each counter is increasing in one section and decreasing in one section only.
Hence, $I_{eq}$ can be partitioned into a finite (but exponentially growing as $\alpha$ and $\beta$ grows) number of subsets, where each is accepted by a machine with $\alpha+\beta$ counters corresponding to
counter $x$, each counter is only increasing (and decreasing) in one section, and the pattern
$C_{(1,x)}, \ldots, C_{(\alpha + \beta,x)}, D_{(\alpha+\beta,x)}, \ldots, D_{(1,x)}$ corresponds to counter $x$.
Proceeding in this fashion for every counter further partitions $I_{eq}$ into more subsets, and one
can see that each instruction language is in $\underline{\StLB}$ as follows: Suppose otherwise.
Let $b_1, \ldots, b_m$ be the sequence of letters in the instruction language of one of the resulting machines. Thus, there exists
$1 \leq l < l' < j < j' \leq m$ such that $b_l = C_s, b_{l'} = C_r, b_j = D_s, b_{j'} = D_r, s \neq r$. Both $C_s$ and $C_r$ could not have been created from the same counter $x$ in $M$, as $x$ created a pattern in
$C_{(1,x)}, \ldots, C_{(\alpha + \beta,x)}, D_{(\alpha+\beta,x)}, \ldots, D_{(1,x)}$. Thus, $C_s$ must have been 
created from $C_x$ of $M$ and $C_r$ must have been created from $C_y$ of $M$, $x \neq y$. But since
each counter used in place of counter $x$ only replaces $C_x$ with some $C_{(1,x)}, \ldots, C_{(\alpha + \beta,x)}$
and $D_x$ with some $D_{(\alpha+\beta,x)}, \ldots, D_{(1,x)}$, and since $I \in \StLB$, we get a contradiction.
As $\LL(\NCM(\underline{\StLB}))$ is closed under union, the result follows.

For $\LB_i\BD_d$, the process is similar to $\LB$ but an extra step is involved.
First, it will be shown that $\LL(\NCM(\LB_i\BD_d)) = \LL(\NCM(\underline{\LB}_i\BD_d))$.
Let $I \in \LB_i \BD_d$. To accept $I_{eq}$ with $\LL(\NCM(\underline{\LB}_i\BD_d))$, the following
idea is used.
First, letters that repeat multiple times
in the {\it letter-bounded increasing} sections are eliminated one at a time
according to a similar procedure.
For example, say $$I = C_1^* C_2^* C_1^* C_3^* C_2^* (D_1 D_2 D_1)^* (D_2 D_3)^* (D_1D_3)^*.$$
Multiple copies of $C_1$ say are eliminated by 
introducing new
counters $(1,1),(1,2),(1,3),(2,1),(2,2),(2,3)$ are introduced, where 
the first coordinate
is over the number of occurrences of $C_1$ in the increasing
section, and the second coordinate is over the number of occurrences
containing $D_1$ in the decreasing section.
Then in $M$, instead of increasing according to pattern 
$C_1$ in the first section, 
$M$ increases counter $(1,1)$, then nondeterministically switches
to $(1,2)$, then to $(1,3)$. Then when simulating the second section of $C_1$, 
$M_1$ increases counter $(2,1)$
then switches to $(2,2)$, then to $(2,3)$. Therefore, $M$
is increasing according to the pattern 
$C_{(1,1)}^* C_{(1,2)}^* C_{(1,3)}^* C_2^* C_{(2,1)}^* C_{(2,2)}^* C_{(2,3)}^* C_3^* C_2^*$. 
In the decreasing section,
$M$ decreases with pattern 
$$(D_{(1,1)}D_2 D_{(1,2)})^* (D_{(1,1)}D_2 D_{(2,2)})^* (D_{(2,1)}D_2D_{(2,2)})^* (D_2 D_3)^* (D_{(1,3)} D_3)^* (D_{(2,3)}D_3)^*.$$ Essentially, $M$ nondeterministically
guesses how much of counter $1$ will be decreased in the various
bounded sections.

After re-numbering these new counters to be consecutive numbers,
and proceeding inductively for every counter where some $C_x$
occurs multiple times, this shows
$\LL(\NCM(\LB_i\BD_d)) = \LL(\NCM(\underline{\LB}_i\BD_d))$.

Next, consider $I = C_{i_1}^* \cdots C_{i_k}^* w_1^* \cdots w_m^* \in \underline{\LB}_i\BD_d$, each 
element of $\Delta_{(k,c)}$ occurs exactly once in 
$C_{i_1}, \ldots, C_{i_k}$, $w_i \in \Delta_{(k,d)}^+$, $1 \leq i \leq m$
and some $D_x$ occurs multiple times in $w_1, \ldots, w_m$.
Multiple copies of $D_x$ will be eliminated for each counter $x$,
one at a time. Then $M$ (the machine where $x$ is removed) is created accepting $I_{eq}$ as follows: 
if $D_x$ occurs $\beta>1$ times (where $D_x$ occurring multiple
times within a single word counts as multiple occurrences)
within the words $w_1, \ldots, w_m$, then introduce new counters
$C_1', \ldots, C_{\beta}'$.
When reading $C_x$,
(which only happens
in one section), $M$ increases counter $C_1'$, then
nondeterministically switches to $C_2'$, etc. until counter
$C_{\beta}'$. Intuitively, $M$ is guessing how much of counter $x$
will be decreased by the $p$th occurrence of $D_x$ in 
$w_1, \ldots, w_m$. Then, when reading $D_x$, $M$ decreases according to 
the pattern $w_1', \ldots, w_m'$, where each $w_i'$ is obtained
from $w_i$ by replacing the $p$th occurrence of $D_x$ with $D_p'$.
($M$ must remember the words $w_1', \ldots, w_m'$ in the finite
control and keep track of which word $w_i'$ and the position
within $w_i'$ it is currently
simulating in order to decrease the counters in the appropriate order.)
For example, if $I$ has $m=1$ and $w_1=(D_1 D_1)^*$, then when increasing according to the pattern $C_1^*$,
$M$ uses counter $C_1'$, then switches to $C_2'$.
Then when decreasing, $M$ decreases according to the pattern
$(D_1'D_2')^*$ (which it can do by remembering
$(D_1' D_2')$ in the finite control). Thus, during the
increase, it is guessing how much will be consumed
by the first and second occurrence of $D_1$ and then
decreasing the appropriate counters.

Proceeding inductively, it follows that, $\LL(\NCM(\LB_i\BD_d)) = \LL(\NCM(\underline{\LB}_i\BD_d)) = \LL(\NCM(\underline{\LB}_i\underline{\BD}_d))$.

The case is similar with $\BD_i\LB_d$.
\qed
\end{proof}

The next goal is to separate some families of $\NCM$ languages that use 
different instruction languages.

A (quite technical) lemma that is akin to a pumping lemma is proven,
but is done entirely on derivations rather than words, so
that it can be used twice starting from the same derivation within
Proposition \ref{witness2}.

First, the following definition is required.
Given an $\NCM$ machine $M$,
a derivation of $M$,
$$(p_0,w_0,c_{0,1}, \ldots, c_{0,k}) \vdash_M^{t_1} \cdots
 \vdash_M^{t_m}(p_m, w_m, c_{m,1}, \ldots, c_{m,k}),$$
is called {\em collapsible}, if there exists $i,j, 0\leq i<j \leq m$ such that
$p_i = p_j, w_i = w_j$, and $c_{i,l} = c_{j,l}$, for all $l$, $1 \leq l \leq k$, and non-collapsible otherwise. It is clear that
given any accepting computation, there is another that can
be constructed that is non-collapsible,
simply by eliminating configurations from the original.

\begin{lemma}
\label{pumpinglemma}
Let $M=(k,Q,\Sigma,\lhd,\delta,q_0,F)$ be a well-formed $k$-counter machine in $\NCM(\underline{\LB})$.
Consider any non-collapsible accepting derivation
$$(p_0,w_0,c_{0,1}, \ldots, c_{0,k}) \vdash_M^{t_1} \cdots
 \vdash_M^{t_m}(p_m, w_m, c_{m,1}, \ldots, c_{m,k}),$$ 
 where $p_0 = q_0, c_{0,j} = c_{m,j} = 0$, for all $1 \leq j \leq k,
 p_m \in F, w_m = \lambda$. Assume that there exists $x, y$, $0 < x \leq y \leq m$
 such that
$p_{x-1} = p_y$, and this state occurs at least $|Q|+2$
times in $p_{x-1}, p_{x}, \ldots, p_y$,
and $|h_{\Sigma}(t_x \cdots t_y)|>0 $,
$h_{\Delta}(t_x \cdots t_y) \in C_i^* \cup D_i^*$, for some $i$, $1\leq i \leq k$.
Then at least one of the following are true:
 \begin{enumerate}
 \item there exists $r,s$ with $x \leq r \leq s \leq y$ and an accepting derivation on transition sequence
 $$t_1 t_2 \cdots t_{r-1} (t_r \cdots t_s)^2 t_{s+1}  \cdots t_m,$$
with $|h_{\Sigma}(t_r \cdots t_s)|>0$, 
\item $h_{\Delta}(t_{x} \cdots t_y) \in C_i^+$ and
there exists $r,s$ with $x \leq y \leq r \leq s$ and $k_1, k_2>1$ 
such that the sequence
$$t_1 t_2 \cdots t_{x-1} (t_{x} t_{x+1} \cdots t_y)^{k_1} t_{y+1} \cdots t_{r-1} (t_r t_{r+1} \cdots t_s)^{k_2} t_{s+1} \cdots t_m,$$ is an accepting computation, and $h_{\Delta}(t_r \cdots t_s) \in D_i^+$,
\item $h_{\Delta}(t_{x} \cdots t_y) \in D_i^+$ and there exists
$r,s$ with $r \leq s \leq x \leq y$ and $k_1,k_2>1$ such that the sequence
$$t_1 t_2 \ldots t_{r-1} (t_r t_{r+1} \cdots t_s)^{k_1} t_{s+1} \cdots t_{x-1} (t_x t_{x+1} \cdots t_y)^{k_2} t_{y+1} \cdots t_m,$$ is an accepting computation, and $h_{\Delta}(t_r \cdots t_s) \in C_i^+$.
\end{enumerate}
\end{lemma}
\begin{proof}
Consider a derivation as above, satisfying the stated assumptions.
Let $q = p_{x-1}=p_y$ be the state, and $i$ the counter.

Between any two consecutive occurrences of $q$ in the
subderivation $t_x, \ldots, t_y$,
if counter $i$ does not change, then at least one input letter
must get read since this derivation is non-collapsible
(and since only counter $i$ can change in this sequence).
Furthermore, repeating this sequence of transitions between $q$ and itself twice
must be an accepting computation, since the state repeats,
the counters have not changed, and at least one extra input letter is read. In this case, item 1 is true.

Otherwise, between every two consecutive occurrences
of $q$ in the subderivation $t_x, \ldots, t_y$,
counter $i$ must change (and either an input letter is read,
or not).
Thus, between the first
and last occurrence of $q$ in $t_x, \ldots, t_y$, at least one input
letter is read (by assumption), and the counter must
change at least $|Q|+1$ times (since $q$ occurs $|Q|+2$ times
in the sequence), either
increasing if $h_{\Delta}(t_x \cdots t_y) \in  C_i^+$, or decreasing if 
$h_{\Delta}(t_x \cdots t_y) \in D_i^+$.

For the first case, assume $h_{\Delta}(t_x \cdots t_y) \in  C_i^+$.
Within $t_x \cdots t_y$, 
counter $i$ increases by $z$ say, where $z>|Q|$.
Hence, counter $i$ must decrease by $z$ as well since $M$
is well-formed. Since the instruction language is in
$\LB$, all the decreasing of counter $i$ must
be within the derivation where no other counter is changed. 
When counter
$i$ decreases, there must also be a state $p$ that appears twice
with at least one decrease in between this repeated state since $z>|Q|$. 
Let $z'>0$ be the amount the counter is decreased between $p$ and
itself.
That is, there must exist $r \leq s$ such that
$p_{r-1} = p = p_s$ and counter $i$ is decreased by $z'>0$ within
this part of the derivation.

Create a new derivation from the derivation above where,
during the cycle that increases counter $i$ by $z$, we increase
by $z + z z'$ (by iterating this cycle $1 + z' = k_1>1$ times), and during the cycle that decreases counter $i$
by $z'$, we instead decrease the counter by $z' + z z'$ (by iterating this cycle $1 + z = k_2>1$ times). This new computation must accept
and item 2 is true. 

The case is similar if $h_{\Delta}(t_x \cdots t_y) \in D_i^+$, with
item 3 being true.
\qed
\end{proof}

The next result follows from Lemma \ref{distinct} and this new pumping lemma.
\begin{proposition} \label{witness}
$\{a^n b^n c^n \mid n >0\} \notin \LL(\NCM(\LB))$.
\end{proposition}
 
\begin{proof}
Assume otherwise. Let $L$ be the language in the statement, and let $M \in \NCM(\underline{\LB})$ accepting $L$, where
each character of $\Delta_k$ occurs exactly once,
$I \subseteq a_1^* \cdots a_{2k}^*$, which is enough by Lemma \ref{LBdistinct}. Let $Q$ be the state set of $M$.

Let $n = (|Q|+1)(|Q|+2)(2k + 1)$. Then, on input $w=a^n b^n c^n$, consider
a non-collapsible accepting computation on
transition sequence $t_1 t_2 \cdots t_m$ of $M$;  
that is,
$$(p_0,w_0,c_{0,1}, \ldots, c_{0,k}) \vdash_M^{t_1} \cdots
 \vdash_M^{t_m}(p_m, w_m, c_{m,1}, \ldots, c_{m,k}),$$ 
 where $p_0 = q_0, c_{0,j} = 0 = c_{m,j}, 1 \leq j \leq k,
 p_m \in F, w_m = \lambda$, and $w = w_0$.

When reading the $a$'s
there must exist $x', y'$, $0 < x' \leq y' \leq m$ such
that $h_{\Delta}(t_{x'} \cdots t_{y'}) \in \{C_i,D_i\}^*$,
for some $i$, $ 1\leq i \leq 2k$,
such that
$|h_{\Sigma}(t_{x'} \cdots t_{y'})| \geq (|Q|+1)(|Q|+2)$
(at least $(|Q|+1)(|Q|+2)$
$a$'s are read while increasing or decreasing counter $i$).
Thus, at least this many transitions are applied during this
sequence of transitions.
Some state $q$ occurs at least $|Q|+2$ times in this subderivation, 
with at least one input letter read between the first and last occurrence of $q$.
Hence, Lemma \ref{pumpinglemma} must apply.

If case 1 is true, this produces a word with more $a$'s than $b$'s.

If case 2 is true, then (using the variables in the Lemma
\ref{pumpinglemma} statement), this derivation has
more than $n$ $a$'s since $k_1>1$ and 
$|h_{\Sigma}(t_x \cdots t_y)|>0$, and therefore
$h_{\Sigma}(t_r \cdots t_s)$ would need to consist
of both $b$'s and $c$'s, otherwise words would be produced
with more $b$'s than $c$'s, or more $c$'s than $b$'s. But then,
there are words that are not in $a^*b^* c^*$, a contradiction.

If case 3 is true, then this produces a word with more $a$'s than $b$'s and $c$'s.
\qed
\end{proof}

In addition, the following can be shown with Lemma \ref{distinct}
and two applications of the pumping lemma.
\begin{proposition} \label{witness2}
$\{a^n b^n c^l d^l \mid n,l>0\} \notin \LL(\NCM(\LB_i\LB_d))$.
\end{proposition}
 
\begin{proof}
Assume otherwise. Let $L$ be the language in the statement, and let $M \in \NCM(\underline{\LB}_i \underline{\LB}_d)$ with $k$ counters accepting $L$ satisfying 
$I \subseteq a_1^* \cdots a_{2k}^*$, which is enough by Lemma \ref{LBdistinct}. Let $Q$ be the state set of $M$.

Let $n = (|Q|+1)(|Q|+2)(2k + 1)$. Then, on input $w=a^n b^n c^n d^n$, consider
a non-collapsible accepting computation on $t_1 t_2 \cdots t_m$ of $M$ accepting $w$;  
that is,
$$(p_0,w_0,c_{0,1}, \ldots, c_{0,k}) \vdash_M^{t_1} \cdots
 \vdash_M^{t_m}(p_m, w_m, c_{m,1}, \ldots, c_{m,k}),$$ 
 where $p_0 = q_0, c_{0,j} = 0 = c_{m,j}, 1 \leq j \leq k,
 p_m \in F, w_m = \lambda$, and $w = w_0$.

When reading the $a$'s
there must exist $x', y'$, $0 < x' \leq y' \leq m$ such
that $h_{\Delta}(t_{x'} \cdots t_{y'}) \in \{C_i,D_i\}^*$,
for some $i$, $ 1\leq i \leq 2k$,
such that
$|h_{\Sigma}(t_{x'} \cdots t_{y'})| \geq (|Q|+1)(|Q|+2)$
(at least $(|Q|+1)(|Q|+2)$
$a$'s are read while increasing or decreasing counter $i$).
Then, at least this many transitions are applied during this
sequence of transitions.
Some state $q$ occurs at least $|Q|+2$ times in this subderivation, 
with at least one input letter read between the first and last occurrence of $q$.
Hence, Lemma \ref{pumpinglemma} must apply.

If case 1 is true, this produces a word with more $a$'s than $b$'s.

If case 3 is true, then this produces a word with more $a$'s than $b$'s.

Assume for the rest of this proof then that case 2 is true.
Then, 
there exists $r,s$ with $x \leq y \leq r \leq s$ and $k_1, k_2>1$ 
such that the sequence
$$t_1 t_2 \cdots t_{x-1} (t_{x} t_{x+1} \cdots t_y)^{k_1} t_{y+1} \cdots t_{r-1} (t_r t_{r+1} \cdots t_s)^{k_2} t_{s+1} \cdots t_m,$$ is an accepting computation, and
$h_{\Delta}(t_r \cdots t_s) \in D_i^+$. The word accepted,
has
more than $n$ $a$'s, $n'$ say, since $k_1>1$ and 
$|h_{\Sigma}(t_x \cdots t_y)|>0$, and the only way
to not obtain a contradiction would be if
$h_{\Sigma}(t_r \cdots t_s)$ consists
of only $b$'s such that the resulting input word reading during this derivation also has $n'$ $b$'s. Also, because 
$h_{\Delta}(t_r \cdots t_s) \in D_i^+$, it follows that
a counter has already started decreasing while reading the $b$'s.
Therefore, after this point of the derivation, no counter
can increase again since $M \in \NCM(\underline{\LB}_i\underline{\LB}_d)$.

Although this new derivation is potentially collapsible (since
new transitions were added in from $t_1 \cdots t_m$), as
mentioned earlier, it is possible to obtain a non-collapsible
derivation from this new derivation simply by removing configurations
(entirely in the new parts added while reading $a$'s and $b$'s).
Then, a new derivation can be obtained on transition sequence
$s_1 \cdots s_{m'}$ accepting $a^{n'} b^{n'} c^n d^n$.

Consider this non-collapsible accepting derivation
and consider the subsequence
when reading the $c$'s.
There must exist $x'', y''$, $0 <  x'' \leq y'' \leq m'$ such
that $h_{\Delta}(t_{x''} \cdots t_{y''}) \in \{D_j\}^*$,
for some $i$, $ 1\leq j \leq k$,
such that
$|h_{\Sigma}(t_{x''} \cdots t_{y''})| \geq (|Q|+1)(|Q|+2)$
(it must be $D_j$ since this derivation has already started
decreasing while reading the $b$'s).
Then, at least this many transitions are applied during this
sequence of transitions.
Some state $q'$ occurs at least $|Q|+2$ times in this subderivation, 
with at least one input letter read between the first and last occurrence of $q'$.
Hence, Lemma \ref{pumpinglemma} must again apply.

If case 1 applies, then this produces a word with more $c$'s than
$d$'s, as does case 3, and case 2 cannot apply since the derivation
has already started decreasing. Thus, we obtain
a contradiction.
\qed
\end{proof}

Therefore, the following is immediate:
\begin{proposition}
\label{separateclasses}
$\LL(\NCM(\LB_i\LB_d)) \subsetneq \LL(\NCM(\LB) )\subsetneq \LL(\NCM(\BD))$.
\end{proposition}
 
\begin{proof}
The inclusions follow from the definitions. Strictness of the first
inclusion follows from Proposition \ref{witness2}, as
$\{a^n b^n c^l d^l \mid n,l>0\} \in  \LL(\NCM(\LB))$ by
making a $2$-counter machine that reads $a$'s and adds the number
of $a$'s to the first counter, then reads $b$'s while verifying
that this number is the same, and then reads $c$'s while adding
the number of $c$'s to the second counter, then reads $d$'s while
decreasing the second counter, verifying that the number is the same.

Strictness of the second inclusion follows from Proposition
\ref{witness}, as $\{a^n b^n c^n \mid n > 0\} \in \LL(\NCM(\BD))$
by building a $2$-counter machine that adds $1$ to counter $1$
then $2$ repeatedly for each $a$ read, then verifies that the contents
of the first counter is the same as the number of $b$'s, then verifies that the contents of the second counter is the same as the number of $c$'s.
\qed
\end{proof}


Next, some of the families will be analyzed individually while creating a more restricted set of generators than what is provided by Proposition \ref{eachSmallest}.

First, two characterizations of $\LL(\NCM(\LB))$ will be given.
Let $k \geq 1$, and let 
$$L_k^{\LB} = \{a_1^{i_1} a_2^{i_2} \cdots a_{2k}^{i_{2k}} \mid \begin{array}[t]{l} \{a_1, \ldots, a_{2k}\} = \Delta_k, \\ \mbox{and~} (C_j = a_l, D_j = a_n \mbox{~implies both~} l < n \mbox{~and~} i_l = i_n), \mbox{~for each~} j, 1 \leq j \leq k\}.\end{array}$$
\begin{proposition}
\label{LBidSmallest}
$\LL(\NCM(\LB))$ is the smallest full trio containing
all distinct-letter-bounded languages of the form
$\{a_1^{i_1} \cdots a_{2k}^{i_{2k}} \mid a_j = C_l, a_n = D_l$ imply
$i_j = i_n\}$, where $\{a_1, \ldots, a_{2k}\} = \Delta_k$ such that $a_j = C_l, a_n = D_l$ implies $j <n$.
Furthermore, $\LL(\NCM(\LB)) = \fulltrio(\{ L_k^{\LB} \mid k \geq 1 \})$.
\end{proposition} 
 
\begin{proof}
For the first part, it follows from Lemma \ref{LBdistinct} and Proposition \ref{generalSmallest} that every language
in $\LL(\NCM(\LB))$ can be obtained by an instruction language in
${\cal I}$, where ${\cal I}$ is the distinct-letter-bounded
subset of $\LB$. Thus, $\LL(\NCM({\cal I})) = \LL(\NCM(\LB))$.
From Proposition \ref{generalSmallest}, it follows that
$\LL(\NCM({\cal I})) = \fulltrio({\cal I}_{eq})$. Furthermore, the languages
in the proposition statement give ${\cal I}_{eq}$ after erasing arbitrary pairs
of $C_i,D_i \in \Delta_k$ with a homomorphism.

For the second part, it is clear that $L_k^{\LB}$ is the finite
union of languages of the form of Proposition \ref{LBidSmallest},
and since this family is closed under union by Proposition \ref{eachSmallest},
then $L_k^{\LB} \in \LL(\NCM(\LB))$.
Further, all bounded languages $I$ of the form of Proposition
\ref{LBidSmallest} can be obtained by intersecting $L_k^{\LB}$ with
the regular language $a_1^* \cdots a_{2k}^*$.
\qed
\end{proof}

Next, a characterization for
$\LL(\NCM(\LB_i\LB_d))$ will be given, whose proof is similar to
Proposition \ref{LBidSmallest}. Let $k \geq 1$, and let 
$$L_k^{\LB_i\LB_d} = \{a_1^{l_1} \cdots a_k^{l_k} b_1^{j_1} \cdots b_k^{j_k} \mid \begin{array}[t]{l} \{a_1, \ldots, a_k \}= \Delta_{(k,c)}, \{b_1,\ldots, b_k\} = \Delta_{(k,d)}, \\ \mbox{and~} (C_m = a_i, D_m = b_n \mbox{~implies~} l_i = j_n), \mbox{~for each~} j, 1 \leq j \leq k\}.\end{array}$$

\begin{proposition}
\label{LBiLBdSmallest}
$\LL(\NCM(\LB_i\LB_d))$ is the smallest full trio containing
all distinct-letter-bounded languages of the form
$\{a_1^{l_1} \cdots a_k^{l_k} b_1^{j_1} \cdots b_k^{j_k} \mid 
a_i = C_m, b_n = D_m$ imply
$l_i = j_n\}$, where $\{a_1, \ldots, a_k\} = \Delta_{(k,c)}$ and $\{b_1, \ldots, b_k\} = \Delta_{(k,d)}$.
Furthermore, $\LL(\NCM(\LB_i\LB_d)) = \fulltrio(L_k^{\LB_i\LB_d} \mid k \geq 1\})$.
\end{proposition}

An interesting alternate characterization will be provided
next for both families using properties of
semilinear sets.
Let $m \ge 1$.
A linear set $Q \subseteq \mathbb{N}_0^n, n \ge 1$,
is {\em $m$-positive} if each periodic vector of $Q$
has at most $m$ non-zero coordinates.
(There is no restriction on the constant
vector.) A semilinear set $Q$ is $m$-positive
if it is a finite union of $m$-positive linear sets.

Let  $L \subseteq a_1^* \cdots a_n^*, a_1, \ldots, a_n \in \Sigma$ be a distinct-letter-bounded
language.
Further, $L$ is called {\em distinct-letter-bounded 2-positive} if there exists a 2-positive semilinear set  $Q$ such that
$L = \{a_1^{i_1} \cdots  a_n^{i_n}  ~|~  (i_1, \ldots, i_n) \in Q \}$.
Further, $L$ is called {\em distinct-letter-bounded 2-positive overlapped} if there exists a 2-positive semilinear set  $Q$
with the property that in any of the linear sets comprising $Q$, there
are no periodic vectors $v$ with non-zero coordinates  at positions
$i  <  j$,  and $v'$ with non-zero coordinates at positions $i' < j'$
such that $1 \le i < j  < i' < j' \le n$, and
$L = \{a_1^{i_1} \cdots a_n^{i_n}  ~|~  (i_1, \ldots, i_n) \in Q \}$.
Hence, they overlap in the sense that, for any such $Q, v, i, j, v', i',j'$,
then the interval $[i,j]$ must overlap with $[i',j']$.

As above, we can also define distinct-letter-bounded 1-positive. Clearly, these languages
are regular and, hence, contained in any nonempty full trio
family \cite{G75}. For 2-positive, the following is true:

\begin{proposition} \label{CHAR}
$~~~$
\begin{enumerate}
\item
$\LL(\NCM(\LB))$ is the smallest full trio containing all languages that are
distinct-letter-bounded 2-positive.
\item
$\LL(\NCM(\LB_i\LB_d))$ is the smallest full trio containing all languages that are
distinct-letter-bounded 2-positive overlapped.
\end{enumerate}
\end{proposition}
 
\begin{proof}
For Part 1, let
$Q \subseteq \mathbb{N}_0^m$ be a 2-positive
semilinear set.  We will show that the language $L = \{a_1^{i_1} \cdots a_m^{i_m}
~|~ (i_1, \ldots, i_m) \in Q\}, a_1, \ldots, a_m \in \Sigma$ is in $\LL(\NCM(\LB))$.
It is sufficient prove the case when $Q$ is a linear set by
Proposition \ref{eachSmallest}.

Consider the constant $c = (c_1, \ldots, c_m)$ and $n$ periods $p_i = (p_{i_1}, \ldots, p_{i_m}), 1 \leq i \leq n$.
Let $f(c) = \{a_1^{c_1} \cdots a_m^{c_m} \}$ and $f(p_i) = \{a_1^{j p_{i_1}} \cdots a_m^{jp_{i_m}} \mid j \geq 0\}$.
Consider new letters $\Sigma' = \{a_{i,j} \mid 0 \leq i \leq n, 1 \leq j \leq m\}$, and
$R = (a_{0,1}^* a_{1,1}^* \cdots a_{n,1}^*) \cdots (a_{0,m}^* a_{1,m}^* \cdots a_{n,m}^*)$.
For $0 \leq i \leq n$, let $h_i$ be a homomorphism that maps $a_{i,j}$ to $a_j$, and erases any $a_{l,j}, l \neq i$,
and let $h$ map any $a_{i,j}$ to $a_j$. Construct an $\NCM(\LB)$ $M$ that, on input $w \in R$ verifies that
$h_i(w) = f(C)$ and $h_i(w) \in f(p_i)$. This is possible as each period only has two non-zero
components. Furthermore, $h(L(M)) = L$, and hence $L \in \LL(\NCM(\LB))$.

Conversely, 
$\LL(\NCM(\LB))$ is the smallest full trio containing
all distinct-letter-bounded languages of the form of Proposition
\ref{LBidSmallest}, by that proposition. Further, all of these are distinct-letter-bounded
$2$-positive. Thus, Part 1 follows.

The proof for Part 2 is similar to the above proof. 
\qed
\end{proof}

Next, a characterization will be given of the smallest
full trio containing all bounded context-free languages.
For that,
we consider instruction family $\StLB$.
An example of an $\StLB$ instruction language (counter behavior)
is  $C_1^* C_2^* C_3^* D_3^* C_2^* D_2^* C_1^* D_1^*$.
On the other hand, behavior
    $C_1^* C_2^* C_3^* D_3^* C_2^* D_2^* C_1^* D_2^* C_1^* D_1^*$
is not an $\StLB$ language since $C_2$ appears, then $C_1$,
then $D_2$, then $D_1$, violating the $\StLB$ definition.

Interestingly, it has previously been found that there does not exist a minimal (non-regular)
full trio with respect to the bounded context-free languages \cite{smallestBounded}.
The next results show that $\LL(\NCM(\StLB))$ is the smallest
full trio containing all bounded context-free languages.
Our proof uses a known
characterization of distinct-letter-bounded
context-free languages ($\CFL$s) from \cite{GinsburgCFLs}.




\begin{proposition}
$\LL(\NCM(\StLB)) = \LL(\NCM(\underline{\StLB})) = \fulltrio(\LL(\CFL)\bd)$.
\end{proposition}
\begin{proof}
The first equality was shown in Lemma \ref{distinct}.
It will be shown that every language in $\LL(\CFL)\bd$ is in
$\LL(\NCM(\StLB))$ from which it will follow that 
$\fulltrio(\LL(\CFL)\bd) \subseteq \LL(\NCM(\StLB))$.
For every $L \in \LL(\CFL)\bd, L \subseteq w_1^* \cdots w_n^*$,
then for distinct $a_1, \ldots, a_n$,
$L' = \{a_1^{i_1} \cdots a_n^{i_n} \mid w_1^{i_1} \cdots w_n^{i_n} \in L\}$ must also be in $\LL(\CFL)$ by using a finite transducer
that reads $w_i$ and outputs $a_i$, as $\LL(\CFL)$ is closed under
finite transductions \cite{G75}. Since $\LL(\NCM(\StLB))$ is
a full trio by Proposition \ref{eachSmallest}, it is sufficient to show that
every  distinct-letter-bounded $\LL(\CFL)$ 
$L' \subseteq a_1^* \cdots a_n^*$ can be
accepted by an $\NCM(\StLB)$, as $L = h(L')$ for a homomorphism
$h$ that outputs $w_i$ from $a_i$.  

Assume that 
$n \ge  2$.  (The case $n =1$ is trivial,
since $L$ is regular.) 

The claim will be proven by induction on $n$. 
The result holds for $n = 2$, since it is known that
for every $\LL(\CFL)$ $L \subseteq w_1^* w_2^*$ (where
$w_1, w_2 \in \Sigma^+)$, $L$ can be accepted by an $\NCM(1,1)$,
hence by an $\NCM(\StLB)$ \cite{IbarraRavikumar} that satisfies
instruction language $C_1^* D_1^*$.

Now suppose $n \ge 3$ and it is true for all values smaller than $n$.
The following characterization is known \cite{GinsburgCFLs}.
For all $\Sigma = \{a_1, \ldots, a_n\}$, $n \ge 3$, then
each $\LL(\CFL)$ $L \subseteq a_1^* \cdots a_n^*$ is a finite union of sets of the
following form:
\begin{center}
$M(D,E, F) = \{a_1^i x y a_n^j ~|~ a_1^i a_n^j \in D, x \in E, y \in F \}$,
\end{center}
where  $D \subseteq a_1^*a_n^*$, $E \subseteq a_1^* \cdots a_q^*$, 
$F \subseteq a_q^* \cdots a_n^*$, $1 < q < n$, are in $\LL(\CFL)$, and
conversely, each finite union of sets of the form $M(D,E,F)$ is a 
$\LL(\CFL)$ $L \subseteq a_1^* \cdots a_n^*$.

By this, $D$ can be accepted by an
$\NCM(\StLB)$ $M_1$ with 1 counter.
By the induction hypothesis, $E$ and $F$ can be
accepted by $\NCM(\StLB)$ machines $M_2$ with $k_2$
and $M_3$ with $k_3$ counters, respectively, since they are over smaller
alphabets.

Since $\LL(\NCM(\StLB))$ is closed under union,
it is sufficient to build a
$\NCM(\StLB)$ machine $M$ accepting $M(D,E,F)$. 
Then $M$ has $k_2 + k_3 +1$ counters.  On a given input, $M$
starts by simulating $M_1$ with counter $1$, and while still reading $a_1$'s, it
remembers the current state of $M_1$ in the finite control,
and starts simulating $M_2$ with counters $2$ through $k_2+1$. (The point when $M$ starts
simulating $M_2$ is nondeterministically chosen, as
long as the input head of $M$ has not gone past
the $a_1$'s.)
After $M_2$ accepts, 
$M$ starts simulating $M_3$ using the remaining counters. (Again, the point when  $M$
starts simulating $M_3$ is nondeterministically chosen.)
When  $M_3$ accepts,
$M$ continues the simulation of $M_1$ from the state it
remembered until the string is accepted.
Hence, $L(M) = M(D,E,F)$, and $M$ satisfies the instruction language $C_1^* D_1^*$ concatenated
with the instruction language of $M_2$ over $\{C_2,D_2,\ldots, C_{k_2+1},D_{k_2+1}\}$, concatenated
with the instruction language of $M_3$ with the remaining letters, followed by $C_1^* D_1^*$. Thus, $M(D,E,F)$, and $L \in \LL(\NCM(\StLB))$.

Conversely, let $I_{eq} \in \underline{\StLB}_{eq}$ over $\Delta_k^*$. So, $I = a_1^* \cdots a_n^*$, where each letter in $\Delta_k$ occurs exactly once in $a_1,\ldots, a_n$, each $C_i$ occurs before $D_i$, and there does not exist $l,l',j,j'$ such that $a_l = C_s, a_{l'} = C_r, a_j = D_s, a_{j'} = D_r$, for some $s \neq r$.
Construct an $\NPDA$  $M$ that accepts $I_{eq}$. 
On input $a_1^{\gamma_1} \cdots a_n^{\gamma_n}$, $M$ pushes all characters in $\Delta_{(k,c)}$, and
on $D_i$, pops $C_i$. Then $M$ accepts at the end of the input if it is at the bottom
of the stack $Z_0$. Let $s$ satisfy
$1 \leq s \leq k$, and let $l$ and $j$ be such that $C_s = a_l$ and $D_s = a_j$ such that
$\gamma_l>0$. Hence $l<j$. When reading $a_j^{\gamma_j}$, then $a_l^{\gamma_l}$ must
be on the pushdown. Assume it is not at the top. Then there exists $C_r, r\neq s$ such
that $C_r$ is at the top. Thus, there exist $l',j'$ with $l < l'< j< j'$ and $C_r = a_l',D_r =a_j'$.
This contradicts that $I \in \underline{\StLB}$. Hence, $I_{eq} \in \LL(\NCM(\underline{\StLB}))$.
\qed
\end{proof}

From this, the following can be determined:
\begin{corollary} \label{cor6}
$~~$
\begin{enumerate}
\item
$\LL(\NCM(\StLB)) \subsetneq \LL(\NCM(\LB))$.
\item
$\LL(\NCM(\StLB)$ and  $\LL(\NCM(\LB_i\LB_d))$ are incomparable.
\end{enumerate}
\end{corollary}
 
\begin{proof}
Obviously, $\LL(\NCM(\StLB)$ is contained in
$\LL(\NCM(\LB))$.
To show proper containment, let
$L = \{a^i b^j c^i d^j ~|~  i , j \ge 1 \}$.
Clearly, $L$ can be accepted by an $\NCM(\LB)$
with counter behavior $C_1^* C_2^*  D_1^* D_2^*$,
but $L$ is not a context-free language. The result follows since
every language in $\LL(\NCM(\StLB)$ is a context-free language.

For Part 2, $L$ is in $\LL(\NCM(\LB_i\LB_d))$
but is not a $\LL(\CFL)$. Now let
$L' = \{a^ib^ic^jd^j ~|~ i, j \ge 1\}$.
$L'$ is a $\LL(\CFL)$, but this is not in  $\LL(\NCM(\LB_i\LB_d))$
by Proposition \ref{witness2}.
\qed
\end{proof}

Next, it will be demonstrated that all bounded semilinear languages
are in two of the language families (and therefore in all
larger families).
\begin{lemma}
All bounded semilinear languages are in $\LL(\NCM(\BD_i \LB_d))$
and in $\LL(\NCM(\LB_i \BD_d))$.
\label{BoundedSemilinearInSubset}
\end{lemma}
 
\begin{proof}
By \cite{CIAA2016}, it is enough to show for all
distinct-letter-bounded semilinear languages. And since
both families are closed under union, by Proposition \ref{eachSmallest}, and 
since every semilinear set is the finite
union of linear sets, it is enough 
to show for all distinct-letter-bounded semilinear languages
induced by a linear set $Q$.

Let $Q \subseteq \mathbb{N}_0^k$ 
be the linear set with 
constant vector $\vec{v_0}$ and periods $\vec{v_1},\ldots, \vec{v_n}$.
A machine $M$ will be created
where input is of the form
$w= a_1^{i_1} \cdots a_k^{i_k}$, each $a_i$ is distinct, and
$M$ accepts $w$ if and only if $(i_1, \ldots, i_k) \in Q$.
We will start with the case for $\BD_i\LB_d$.

First, $M$ adds $v_0(i)$ to counter $i$, for 
each $i$ from $1 \leq i \leq k$, on $\lambda$ transitions. 
Then $M$ does the same for $\vec{v_1}$ arbitrarily many times, then the same for $\vec{v_2}$, etc. 
This insertion pattern is in the bounded language
$$C_1^* \cdots C_k^* (C_1^{\vec{v_1}(1)} C_2^{\vec{v_1}(2)} \cdots C_k^{\vec{v_1}(k)})^* \cdots
(C_1^{\vec{v_n}(1)} C_2^{\vec{v_n}(2)} \cdots C_k^{\vec{v_n}(k)})^*.$$ The counter contents are in the linear set without having read any input. Then it verifies that the input is equal to the counter contents one counter at a time, and therefore $M$ accepts the distinct-letter-bounded language induced by $Q$. 
So the decreasing pattern is $D_1^* D_2^* \cdots D_k^*$ which is
letter-bounded.

For $\LB_i \BD_d$, this pattern is inverted, and $M$
starts by placing $i_j$ in counter $j$, for $1 \leq j \leq k$,
according to the letter-bounded pattern $C_1^* \cdots C_k^*$.
Then, $M$ subtracts from the counters with the same pattern
that it increased from the counters in the case above, which
is in the bounded language (replacing all $C$'s with $D$'s).
$M$ accepts if all counters reach zero in this fashion. 
\qed
\end{proof}

From the definition, it is immediate that if ${\cal I} \subseteq {\cal I}'$, then $\LL(\NCM({\cal I})) \subseteq \LL(\NCM({\cal I}'))$. 
It is clear that all of $\LB, \BD_i\LB_d, \LB_i\BD_d$ are a subset of $\BD$.
It will be shown that three of these counter families coincide.

\begin{proposition} \label{equal}
$\LL(\NCM(\BD_i\LB_i)) = \LL(\NCM(\LB_i\BD_d)) = \LL(\NCM(\BD))$
is the smallest full trio containing all bounded 
semilinear languages, and the smallest full trio containing
all bounded Parikh semilinear languages.
\end{proposition}
 
\begin{proof}
All of the families are full trios by Proposition \ref{semiAFL}. 
All must contain all bounded semilinear languages by 
Lemma \ref{BoundedSemilinearInSubset}, and therefore all
bounded Parikh semilinear languages \cite{CIAA2016}.

To complete the proof, we will now show that all languages  
in $\LL(\NCM(\BD))$ can be obtained from
the bounded Parikh (which are then bounded Ginsburg) 
semilinear languages using full trio operations.

Let ${\cal I} = \BD$.
By Proposition \ref{generalSmallest}, $\LL(\NCM(\BD))$ is the
smallest family of languages containing
${\cal I}_{eq} = 
\{ I \mid I = \{w=w_1^* \cdots w_m^* \mid |w|_{C_i} = |w|_{D_i}, \mbox{~for each~} 1 \leq i \leq k, \mbox{~all $C_i$'s appear before any $D_i$'s} \}, w_i \in \Delta_k^*, k \geq 1\}$.
But every $I_{eq} \in {\cal I}_{eq}$ is a bounded Parikh semilinear language. Thus the statement follows.
\qed
\end{proof}

\begin{corollary}
For all ${\cal I} \in \{\BD_i\LB_d, \LB_i\BD_d, \BD, \LB_d, \LB_i,\LB_{\cup},\ALL \}$, the family $\LL(\NCM({\cal I}))$ contains all bounded
semilinear languages, and all bounded languages
in $\LL(\NCM)$.
\end{corollary}

Next,  two simple sets of generators for $\LL(\NCM(\BD))$ are established. These languages will therefore be a simple mechanism to
show whether or not a full trio $\LL$ contains every bounded Ginsburg semilinear language, and therefore has exactly the same bounded
languages as $\NCM$, and has all bounded languages contained in
any semilinear trio. 
Let $k \geq 1$, and let $$L_k^{\BD_i\LB_d} = \{w_1^{x_1} \cdots w_m^{x_m} D_1^{y_1} \cdots D_k^{y_k} \mid \begin{array}[t]{l} w_j \in \Delta_{(k,c)}^+,
x_j>0, 1 \leq j \leq m, \mbox{~for~} 1 \leq i \leq k, |w_1 w_2 \cdots w_m|_{C_i} = 1, \\ ( C_i \in \alp(w_j) \mbox{~implies~} y_i = x_j) \},\end{array}$$ 
$$L_k^{\LB_i\BD_d} = \{C_1^{y_1} \cdots C_k^{y_k} w_1^{x_1} \cdots w_m^{x_m}  \mid \begin{array}[t]{l} w_j \in \Delta_{(k,d)}^+,
x_j>0, 1 \leq j \leq m, \mbox{~for~} 1 \leq i \leq k, |w_1 w_2 \cdots w_m|_{D_i} = 1, \\ ( D_i \in \alp(w_j) \mbox{~implies~} y_i = x_j) \}.\end{array}$$
\begin{proposition}
\label{generatorboundedginsburg}
$$\LL(\NCM(\BD_i\LB_d)) = \LL(\NCM(\LB_i\BD_d)) = \LL(\NCM(\BD)) = \fulltrio(L_k^{\BD_i\LB_d} \mid k\geq 1\}) = 
\fulltrio(\{ L_k^{\LB_i\BD_d} \mid k \geq 1\}).$$
\end{proposition}
 
\begin{proof}
The equality of the first three families follows from Proposition \ref{equal}.
First, consider $L_k^{\LB_i\BD_d}$. To show
this language is in $\LL(\NCM(\LB_i\BD_d))$, notice
that $L_k^{\LB_i\BD_d} \cap C_1^* \cdots C_k^* w_1^* \cdots w_m^*$
is clearly in this family for fixed words $w_1, \ldots, w_m$, where
$|w_1 \cdots w_m|_{D_i} = 1$ for each $i$. Furthermore, there
are a finite number of combinations of such words, and
$\LL(\NCM(\LB_i\BD_d))$ is closed under union by Proposition 
\ref{eachSmallest}, therefore $L_k^{\LB_i\BD_d}$ is in this family.

From Lemma \ref{distinct} and Proposition \ref{generalSmallest}, it follows that
$\LL(\NCM(\LB_i\BD_d)) = \fulltrio((\underline{\LB}_i \underline{\BD}_d)_{eq})$.

First consider a slight variant of $L_k^{\LB_i\BD_d}$,
$$L_k = \{C_{i_1}^{y_1} \cdots C_{i_k}^{y_k} w_1^{x_1} \cdots w_m^{x_m}  \mid \begin{array}[t]{l} w_j \in \Delta_{(k,d)}^+,
x_j>0, 1 \leq j \leq m, C_{i_1}, \ldots , C_{i_k} \mbox{~is a permutation of~} \Delta_{(k,c)},\\ \mbox{for~} 1 \leq l \leq k, |w_1 w_2 \cdots w_m|_{D_l} = 1, ( D_{i_l} \in \alp(w_j) \mbox{~implies~} y_l = x_j) \}.\end{array}$$

Every language $I \in (\underline{\LB}_i \underline{\BD}_d)_{eq}$ over
$\Delta_k^*$ for some $k \geq 1$ is equal to 
$\{C_{i_1}^{y_1} \cdots C_{i_k}^{y_k} w_1^{x_1} \cdots w_m^{x_m} \mid x_j = y_l$ if $D_{i_l} \in \alp(w_j)>0
\}$, 
where $C_{i_1}, \ldots, C_{i_k}$ is a permutation of
$\Delta_{(k,c)}$, $w_j \in \Delta_{(k,d)}^+$, and each letter
of $\Delta_{(k,d)}$ occurs exactly once in $w_1 w_2 \cdots w_m$. 
In this case, $I = L_k  \cap C_{i_1}^+ C_{i_2}^+ \cdots C_{i_k}^+ w_1^+ \cdots w_m^+$.
Furthermore, consider the homomorphism $h$ that maps $C_{i_l}$ to $C_{l}$, and $D_{i_l}$ to $D_{l}$.
Then, $I = h^{-1}(L_k^{\BD_i\LB_d} \cap C_1^+ C_2 ^+ \cdots C_k^+ h(w_1)^+ \cdots h(w_m)^+)$. 

The case is similar for $L_k^{\BD_i\LB_d}$.
\qed
\end{proof}

By Proposition \ref{equal}, Proposition \ref{generatorboundedginsburg}, and \cite{CIAA2016}, the following is true:
\begin{proposition}
\label{fullcharacterization}
Let $\LL$ be a full trio. The following are equivalent:
\begin{itemize} 
\item $\LL$ contains all bounded
semilinear languages,
\item $\LL$ contains all distinct-letter-bounded 
semilinear languages,
\item $\LL$ contains all bounded Parikh semilinear languages,
\item $\LL(\NCM)\bd (= \LL(\DCM)\bd = \LL(\NCM(\BD))\bd)$ is contained in $\LL$,
\item $\LL(\NCM(\BD_i \LB_d)) (= \LL(\NCM(\LB_i\BD_d)) = \LL(\NCM(\BD)))$ is contained in $\LL$,
\item $\LL$ contains $L_k^{\BD_i\LB_d}$, for each $k\geq 1$,
\item $\LL$ contains $L_k^{\LB_i\BD_d}$, for each $k \geq 1$.
\end{itemize}
\noindent
Furthermore, if $\LL$ is also semilinear,
then
these conditions are equivalent to
$\LL\bd = \LL(\NCM)\bd =\LL(\DCM)^{bd}$.
\end{proposition}

By Proposition \ref{witness} and Proposition \ref{fullcharacterization},
the following is immediate:
\begin{corollary}
$\LL(\NCM(\LB))$ and $\LL(\NCM(\LB_i\LB_d))$ do not contain all
bounded semilinear languages, or all
bounded Parikh semilinear languages.
\end{corollary}

%
There is another simple equivalent form of the family $\LL(\NCM(\BD))$.
Let $\SBD$ be the subset of $\BD$ that is the family
$$\{I \mid k \geq 1, I = w_1^* \cdots w_m^*,
w_i \in \Delta_k^+, |w_i| \leq 2, 1 \leq i \leq m \}.$$

\begin{proposition} \label{sbd}
$\LL(\NCM(\SBD)) = \LL(\NCM(\BD))$.
\end{proposition}
\begin{proof}
Let $k \geq 1$ and consider $$L_k^{\BD_i\LB_d} = \{w_1^{x_1} \cdots w_m^{x_m} D_1^{y_1} \cdots D_k^{y_k} \mid \begin{array}[t]{l} w_j \in \Delta_{(k,c)}^+,
x_j>0, 1 \leq j \leq m, \mbox{~for~} 1 \leq i \leq k, |w_1 w_2 \cdots w_m|_{C_i} = 1, \\ ( C_i \in \alp(w_j) \mbox{~implies~} y_i = x_j) \},\end{array}$$ 
from Proposition \ref{generatorboundedginsburg}. Construct $M$ to be
a $k$-counter machine that reads $w_1^{x_1} \cdots w_m^{x_m}$ from left-to-right. It can determine the individual words $w_i$ based on repeat patterns and remember them in the finite control (there are $m \leq k$ of them) while adding $x_i$ to counter $i$. Then, it reads $D_1^{y_1}$ and let $i$ be such that $C_1$ is in $w_i$. For every
$D_1$ read, it decreases counter $i$ by one, and if $w_1$ has more than one letter, it increases the next unused counter by one. This has the effect of making a copy of counter $i$, while verifying that $x_i = y_1$. This process can clearly continue with a total of $k$ counters. The instruction language is in $C_1^* \cdots C_m^*$ followed by $(D_iC_{m+1})^*$, etc.
\qed
\end{proof}

Thus, $\SBD$
is enough to generate all bounded Ginsburg semilinear languages,
whereas $\LB$ is not.

Next, we will explore the language families
$\NCM(\LB_d)$ and $\NCM(\LB_i)$. Let $k \geq 1$, and let
$$L_k^{\LB_d} = \{ w_0 w_1 \cdots w_k \mid \begin{array}[t]{l} w_i \in \{C_{i+1},  C_{i+2},\ldots, C_k, D_i\}^*, 0 \leq i \leq k,\\ |w_0 w_1 \cdots w_{j-1}|_{C_j} = |w_j|_{D_j} >0, 1 \leq j \leq k\} \subseteq \Delta_k^*,\end{array}$$
where we define $w_0 \in \Delta_{(k,c)}^*$, and $w_k \in D_k^*$.
\begin{proposition}
$\LL(\NCM(\LB_d)) = \fulltrio(\{L_k^{\LB_d} \mid k \geq 1 \})$.
\label{smallestLBd}
\end{proposition}
 
\begin{proof}
First, $L_k^{\LB_d}$ can be accepted by $M \in \NCM(\LB_d)$
by guessing a partition into $w_0 w_1 \cdots w_k$, and
while reading $w_i$, verify it is in $\{C_{i+1},  C_{i+2},\ldots, C_k, D_i\}^*$, incrementing counter $j$ for every $C_j$ read,
and decrementing counter $i$ for every $D_i$ read, finishing with
all counters empty.

From Lemma \ref{LBdistinct} and Proposition \ref{generalSmallest},
it follows that $\LL(\NCM(\LB_d))$ is the smallest full trio containing all 
$I \subseteq \Delta_k^*, k \geq 1$ such that
$$I = \{ w \mid \begin{array}[t]{l} w \in Y \shuffle Z, Y = \Delta_{(k,c)}^*, 
Z = D_{i_1}^* \cdots D_{i_k}^*, D_{i_1}, \ldots, D_{i_k} \mbox{~is
a permutation}\\ \mbox{of~} \Delta_{(k,d)}, |w|_{C_i}= |w|_{D_i}>0, \mbox{~all~} \mbox{$C_i$'s appear before any $D_i$'s},
\mbox{for~}1 \leq i \leq k \}.\end{array}$$
Notice, every such $I$ can be obtained from
$$I' = \{ w \mid \begin{array}[t]{l} w \in Y \shuffle Z, Y = \Delta_{(k,c)}^*, 
Z = D_{1}^* \cdots D_{k}^*, |w|_{C_i}= |w|_{D_i}>0, \mbox{~all}\\ \mbox{$C_i$'s appear before any $D_i$'s, for~}1 \leq i \leq k \}.\end{array}$$ via homomorphisms that permute elements
of $\Delta_{(k,d)}$ such that $h(D_i) = D_j$ implies $h(C_i) = C_j$.

Notice that there is only one such $I'$ for each $k$,
and it is equal to $L_k^{\LB_d}$.
\qed
\end{proof}

The next proposition follows with a similar proof. Let $k \geq 1$, and let
$$L_k^{\LB_i} = \{ w_0 \cdots w_k \mid \begin{array}[t]{l} w_i \in \{D_1, \ldots, D_{i}, C_{i+1}\}^*, 0 \leq i \leq k,\\ |w_{j-1}|_{C_j} = |w_j w_{j+1} \cdots w_k|_{D_j} >0, 1 \leq j \leq k\} \subseteq \Delta_k^*,\end{array}$$
where $w_0 \in C_1^*$, and $w_k \in \Delta_{(k,d)}^*$.
\begin{proposition}
$\LL(\NCM(\LB_i)) = \fulltrio(\{ L_k^{\LB_i} \mid k \geq 1\})$.
\end{proposition}

Later, in Corollary \ref{reverse}, it will be seen that 
$L \in \LL(\NCM(\LB_d))$ if and only if $L^R \in \LL(\NCM(\LB_i))$.

\begin{figure}[htbp!]
\begin{center}
\includegraphics[width=3.2in]{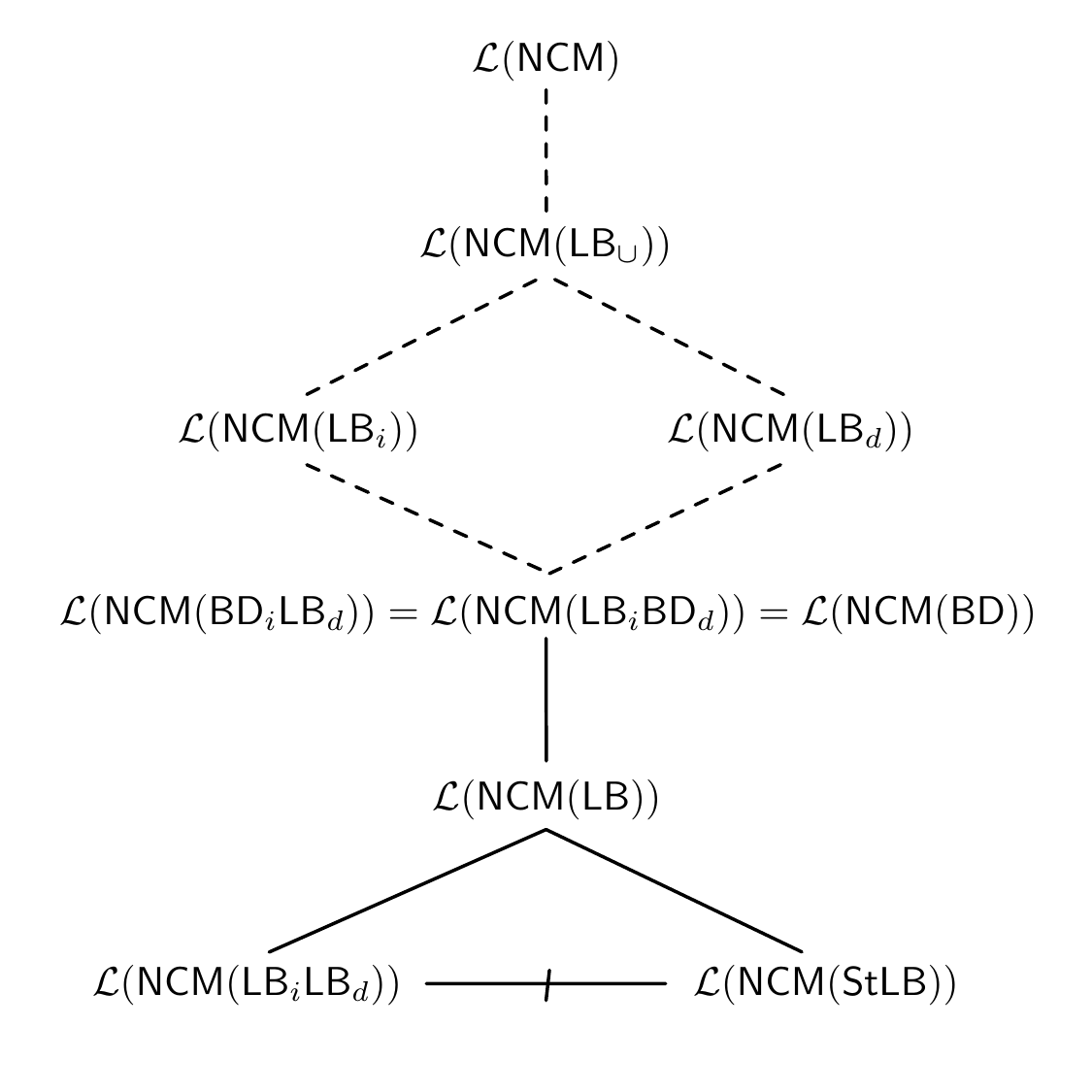}
\caption{Pictured above is a comparison of the computational capacities of the various subfamilies of $\LL(\NCM)$ based on varying the instruction language families. Solid lines indicate strict containments, dotted lines indicate containment where strictness is open, the solid line with the slash indicates that the families are incomparable, and it is open whether the two families where no lines occur between them are incomparable.}
\end{center}
\label{comparison}
\end{figure}

Hence, in this section many families of instruction languages were studied, along with the $\NCM$ machines using these families. A comparison of the computational power of the various $\NCM$ subfamilies is visualized in Figure \ref{comparison}.

\section{Closure Properties}

In this section, some closure properties of the subfamilies of $\NCM$ will be discussed.

\begin{proposition} \label{closure1}
$~~$
\begin{enumerate}
\item
$\LL(\NCM(\LB))$, $\LL(\NCM(\StLB))$, $\LL(\NCM(\BD))$,
$\LL(\NCM(\LB_d))$, $\LL(\NCM(\LB_i))$, $\LL(\NCM(\LB_{\cup}))$ are closed under
concatenation.
\item $\LL(\NCM(\LB_i\LB_d))$ is not closed under concatenation.
\item
$\LL(\NCM(\LB)), \LL(\NCM(\StLB)),$ and $\LL(\NCM(\LB_i\LB_d))$ are not closed under intersection or shuffle.
\item
$\LL(\NCM(\LB))$, $\LL(\NCM(\BD))$, and $\LL(\NCM(\LB_{\cup}))$
are closed under reversal.
\item For each ${\cal I}$ in Definition
\ref{def:families}, $\LL(\NCM({\cal I}))$ is
not closed under complementation.
\item For each ${\cal I}$ in Definition
\ref{def:families}, $\LL(\NCM({\cal I}))$ is
not closed under Kleene $*$.
\end{enumerate}
\end{proposition}
\begin{proof}
$~~$
\begin{enumerate}
\item
Let $M_1$ and $M_2$ be in $\NCM(\LB)$  
with $k_1$ and $k_2$ 1-reversal counters, respectively.
Clearly, a machine $M$ in
$\NCM(\LB)$ can be constructed (resp., $\NCM(\BD)$) with $k_1+k_2$
1-reversal counters to accept $L(M_1)L(M_2)$. Similarly with all other
cases.
\item This follows from Proposition \ref{witness2}.
\item
It is immediate that all are not closed under intersection
or shuffle by Corollary \ref{smallestIntersection}, Proposition \ref{separateclasses}, and Corollary \ref{cor6}.
\item
For reversal, let $M$ be in $\NCM(\LB)$
(resp., $\NCM(\BD)$, $\NCM(\LB_{\cup})$) with $k$ counters over input
alphabet $\Sigma$.  Assume that $M$ accepts in a unique
accepting state with all counters zero.
Construct an NFA $M_1$
with input alphabet  $\Sigma \cup \Delta_k$.
$M_1$ simulates  the computation of $M$ on input $w \in \Sigma^*$,
but whenever $M$ increments (resp., decrements) counter $i$,
$M_1$ reads $C_i$ (resp., $D_i$) instead on the input.
$M_1$ accepts if and only if $M$ accepts. Let $M_2$     
be an NFA accepting $(L(M_1))^R$. Now, construct from $M_2$ an
$\NCM$ $M'$ over input alphabet $\Sigma \cup \Delta_k$
which simulates $M_2$ but whenever $M'$ sees a $D_i$
(resp., $C_i$), it increments (resp., decrements) counter $i$.
$M'$ accepts if $M_2$ accepts and all counters are zero.  
Clearly, $M'$ is in $\NCM(\LB)$
(resp., $\NCM(\BD)$, $\NCM(\LB_{\cup})$).
Let $h$ be a homomorphism defined by $h(a)= a$ for
each $a \in \Sigma$ and $h(C_i) = h(D_i) = \lambda$
for each $i$.  Clearly, $h(L(M'))= (L(M))^R$.
The result follows since these families are full trios;
hence, are closed under homomorphism.
\item Let $L = \{x \# y \mid x,y \in \{a,b\}^+, x \neq y \}$.
Clearly $L$ can be accepted by a well-formed $\NCM(1,1)$ machine which is in all
the families.  We claim that $\overline{L}$ is not in $\LL(\NCM)$.  Otherwise,
$\overline{L} \cap \{a,b\}^* \# \{a,b\}^* = \{ x \#y  ~|~ x, y \in  \{a,b\}^+, x = y \}$
would also be in $\LL(\NCM)$. But it is known that the smallest intersection-closed
full trio (which $\NCM$ is \cite{Ibarra1978}) containing $L$ is the family
of recursively enumerable languages \cite{ResetMachines}.
\item Let $L = \{a^n b^n ~|~ n \ge 1\}$. $L$ can be accepted by all variants. But it was shown in \cite{G78}
that $L^*$ is not in $\LL(\NCM)$.
\qed
\end{enumerate}
\end{proof}

The proof of Part 4 in Proposition \ref{closure1}
also applies to showing:

\begin{corollary} \label{reverse}
$L \in \LL(\NCM(\LB_d))$ if and only if $L^R \in \LL(\NCM(\LB_i))$,
and $L \in \LL(\NCM(\LB_i))$ if and only if $L^R \in \LL(\NCM(\LB_d))$.
\end{corollary}

There are still several open problems.\\
\noindent
{\bf Conjectures}.  We have the following conjectures
concerning some closure properties:
\begin{enumerate}
\item
Let $L = \{w \#a^i b^j \mid w \in \{a,b\}^*, |w|_a = i, |w|_b = j\}$.
Then $L$ is in $\LL(\NCM(\LB_d))$, but we conjecture that $L$ is not in 
$\LL(\NCM(\LB_i))$. If this is correct, it would imply that
$L^R$ is in $\LL(\NCM(\LB_i))- \LL(\NCM(\LB_d))$ because $L^R$ is in
$\LL(\NCM(\LB_i))$, and if it was in $\LL(\NCM(\LB_d))$, then $L^{RR} = L$
would be in $\LL(\NCM(\LB_i))$. It would also imply
$\LL(\NCM(\BD)) \subsetneq \LL(\NCM(\LB_d)) \subsetneq
\LL(\NCM(\LB_{\cup}), \LL(\NCM(\BD)) \subsetneq \LL(\NCM(\LB_i)) \subsetneq \LL(\NCM(\LB_{\cup}))$, and 
$\LL(\NCM(\BD)), \LL(\NCM(\LB_i)), \LL(\NCM(\LB_d))$ are not closed under intersection or shuffle.

\item
$\LL(\NCM) - \LL(\NCM(\LB_{\cup})) \ne \emptyset$.
The candidate witness language is
$L = \{ w~|~ w \in \{a,b\}^+, |w|_a = |w|_b > 0 \}$.
$L$ can be accepted by an $\NCM$ (see Example \ref{ex4})
but does not seem to be in $\LL(\NCM(\LB_{\cup}))$.
If the above item is true, then $\LL(\NCM(\LB_{\cup}))$ is not
closed under intersection since $\LL(\NCM)$ is the smallest
intersection closed full trio containing $\LL(\NCM(\LB_i \LB_d))$.

\end{enumerate}

\section{Applications To Existing Families}

In this section, the results obtained thus far in this paper will be applied to quickly
characterize the bounded languages inside known language families.
First, it was recently shown that the family of finite-index $\ETOL$ languages contains all bounded semilinear languages, and therefore its bounded languages coincide with those in $\LL(\DCM)$ \cite{CIAA2016}.
This could also be shown by proving that $\LL(\NCM(\BD))$ is contained in finite-index $\ETOL$ languages, although we will not
reprove this here.

Next, let $\TMFC$ be the class of
Turing machines with a one-way read-only input tape, and a
finite-crossing\footnote{There is a fixed $c$ such that the number of times the boundary between any two adjacent input
cells is crossed is at most $c$.} read/write worktape. 
Similarly, let $\TMRB$ be the class of Turing machines with a one-way read-only input tape, and a 
reversal-bounded\footnote{There is a fixed $c$ such that the number of times the read/write head switches from moving left to right and vice versa is at most $c$.} read/write worktape. Clearly,
$\LL(\TMRB) \subseteq \LL(\TMFC)$.
Both language families are semilinear
full trios \cite{Harju2002278}. Therefore, $\LL(\TMFC)\bd \subseteq \LL(\NCM)\bd$.
To show that there is equality, 
a simulation of $\NCM(\BD_i\LB_d)$ will be done. Let $M$ be a well-formed
$k$-counter machine satisfying instruction language
$I \subseteq w_1^* w_2^* \cdots w_l^* D_{i_1}^* \cdots D_{i_n}^*$,
where $w_i \in \Delta_{(k,c)}^*, 1 \leq i \leq l, D_j \in \Delta_{(k,d)},
1 \leq j \leq n$. Build a
$\TMRB$ machine $M'$ with worktape alphabet $\Delta_k$ that, 
on input $w$, simulates a
derivation of $M$, whereby, if $M$ increases from counters in the
sequence $C_{j_1}, \ldots, C_{j_m}$, $M'$ instead writes this
sequence on the worktape. Then, $M'$ simulates the decreasing
transitions of $M$ as follows: for every section of decreases in
$D_{i_j}^*$, for $1 \leq j \leq n$, $M'$ sweeps the worktape from
right-to-left, and corresponding to every decrease, replaces
the next $C_{i_j}$ symbol with the symbol $D_{i_j}$ (thereby marking the symbol). This requires $n$ sweeps of the worktape, and $M'$ accepts
if all symbols end up marked and the simulated computation is in a final state.
\begin{proposition}
\label{TCAproof}
$\LL(\NCM(\BD_i\LB_d)) \subseteq \LL(\TMRB)$ and
$\LL(\TMRB)\bd = \LL(\TMFC)\bd = \LL(\DCM)\bd$.
\end{proposition}

Next, the family of multi-push-down automata and languages introduced in \cite{multipushdown} will be discussed. 
Let $\MP$ be these machines. 
They have some number $k$ of pushdowns, and allow to push to every pushdown, but only pop from the first non-empty pushdown. This can clearly simulate every machine in $\NCM(\underline{\LB}_d)$ 
(distinct-letter-bounded, which is enough to accept every language in
$\LL(\NCM(\LB_d))$ by Lemma \ref{distinct}). Furthermore, it follows from results within \cite{multipushdown} that $\LL(\MP)$
is closed under reversal (since it is closed under homomorphic
replication with reversal, and homomorphism). Therefore,
$\LL(\MP)$ also contains $\LL(\NCM(\LB_{\cup}))$.  Also, this family
only contains semilinear languages \cite{multipushdown}.
Hence, the bounded languages within $\LL(\MP)$ coincide with
those in $\LL(\NCM)$ and $\LL(\DCM)$.
\begin{proposition}
\label{MPproof}
$\LL(\NCM(\LB_{\cup})) \subseteq \LL(\MP)$ and 
$\LL(\MP)\bd = \LL(\DCM)\bd$.
\end{proposition}

The simplicity of these proofs using automata that behave quite differently from $\NCM$s, validates the use of instruction languages and subfamilies of $\LL(\NCM)$.

Finally, it is worth noting that it was shown that the smallest intersection-closed full trio containing the
language $\{a^n b^m c^{nm} \mid m,n \in \mathbb{N}\}$ contains all recursively enumerable
bounded languages \cite{boundedAFLs}.

\section{Decidable and Undecidable Properties}
\label{decproperties}

In this section, some decision problems concerning the
$\NCM$ instruction language families are examined.
First, decidability of testing whether a well-formed
counter machine satisfies
some instruction languages will be addressed.

The following results from \cite{Ibarra1978} are required.
An $\NPCM$ is a one-way $\NCM$ augmented with an unrestricted pushdown stack.
\begin{proposition} \cite{Ibarra1978} \label{known}
The following are true: 
\begin{enumerate}
\item
The emptiness problem (``Is the language accepted empty?'') for
$\NPCM$ (hence, also for $\NCM$) is decidable.
\item
The infiniteness problem (``Is the language accepted infinite?'') for
$\NPCM$ (hence, also for $\NCM$) is decidable.
\item
If $M_1$ is an $\NPCM$ (resp., $\NCM$) and $M_2$ is an $\NCM$,
then we can effectively construct an $\NPCM$ (resp., $\NCM$)
$M$ to accept $L(M_1) \cap L(M_2)$.  
\item
It is decidable, given an $\NPCM$ $M_1$ and an $\DFA$ $M_2$,
whether $L(M_1) \subseteq L(M_2)$  (this follows from (1) and (3)).
\end{enumerate}
\end{proposition} 
For this proposition, $\DPCM$ is the deterministic
version of $\NPCM$.
\begin{proposition}
\label{fixedI}
It is decidable, given a well-formed $k$-counter $\NCM$ machine $M$,
and a regular instruction language $I \subseteq \Delta_k^*$ (or even an $\LL(\DPCM)$ instruction language),
whether $M$ satisfies $I$. Hence, it is decidable, for all instruction languages $I$ in any family of
Definition \ref{def:families} and well-formed $M$, whether $M$ satisfies $I$.
\end{proposition}
\begin{proof}
From $M$, another well-formed $k$-counter machine 
$M'$ is built over
$\Delta_k = \{C_1, D_1, \ldots, C_k, D_k\}$.
$M'$ on input $x \in \Delta_k^*$ simulates the
computation  $M$ on some input $w$ (guessing the symbols
comprising $w$) and checking that the counter behavior
when $M$ accepts is $x$.

Construct an $\NPCM$ $M''$ that accepts 
$L(M') \cap \overline{I}$.
This is in $\NPCM$ since $\DPCM$ is
closed under complement \cite{Ibarra1978}, and the intersection of an
$\NPCM$ language with an $\NCM$ language is a $\NPCM$ language by Proposition \ref{known} (3).
Then $M$ does not satisfy $I$ if and only if there 
exists $w \in L(M)$ and a derivation of $w$ 
that uses a sequence of instructions $\alpha$
that is not in $I$ if and only if $L(M') \cap \overline{I} \neq \emptyset$.
Then, only the testing emptiness of $L(M'')$ is needed, which is decidable by Proposition \ref{known} (1).
\qed
\end{proof}
However, note that this
decidability fixes the instruction language $I$ in advance.
This does not imply that for the families ${\cal I}$ of
Definition \ref{def:families}, it is decidable whether or not
a well-formed counter machine $M$ satisfies {\bf some}
$I \in {\cal I}$.
This question, where only the family ${\cal I}$ is given (and not the specific $I \in {\cal I}$),
will be addressed next.

Let $m \ge 1$.
A language $L$ is $m$-bounded if
$L \subseteq w_1^* \cdots w_k^*$ for some
(not-necessarily distinct) 
words $w_1, \ldots, w_k$ such
that $|w_i| =  m$ for each $1 \le i \le k$. Note
that 1-bounded is the same as letter-bounded.

The following lemma was given in \cite{DLT2015journal}.
For completeness, a proof is given while including a few details
missing in \cite{DLT2015journal},
since this result is needed in the proof of
Lemma \ref{decide1}.

\begin{lemma} \label{1-bounded-1}
It is decidable, given an $\NPCM$ $M$, whether $L(M)$ is letter-bounded. If it is letter-bounded,
we can effectively find a $k$ and not-necessarily
distinct symbols $a_1, \ldots, a_k$  such that
$L(M) \subseteq  a_1^* \cdots a_k^*$.
\end{lemma}
\begin{proof}
Construct from $M$ an $\NPCM$ $M'$ accepting a unary language that is a subset of $1^*$.  $M'$ has one more counter, $C$, than $M$.
Then $M'$, on input $1^k, k \ge 1$, simulates the computation
of $M$ on some input $w$ by guessing the symbols of $w$
symbol-by-symbol.  During the simulation, $M'$ increments the counter $C$  if the next input symbol it guesses is different
from the previous symbol it guessed.  When $M$ accepts,
$M'$ checks and accepts if the value of the counter $C$ is at least $k$.   Clearly, $M$ is not letter-bounded if and only if
$L(M')$ is infinite, which is decidable since the infiniteness
problem for $\NPCM$s is decidable by Proposition \ref{known} Part 2.

Part 2 follows from Part 1 by exhaustive search:
we systematically try all sequences $a_1, \ldots, a_k$
(each $a_i \in \Sigma$, starting with $k = 1$)
and check if  $L(M) \subseteq R$, where $R = a_1^* \cdots a_k^*$
(this is decidable by Proposition \ref{known}, part 4
since $R$ is regular).
\qed
\end{proof}

This helps to show the following regarding $\NCM$s.
\begin{proposition}
It is decidable, given a well-formed $k$-counter $\NCM$ machine $M$, whether
$M \in \NCM(\LB_d)$ (resp., $\NCM(\LB_i)$, $\NCM(\LB)$, $\NCM(\LB_i\LB_d)$, $\NCM(\StLB))$).
\end{proposition}
\begin{proof}
Construct a $k$-counter $\NCM$ $M'$ which has 
input alphabet $\Delta_{(k,d)} = \{D_1, \ldots, D_k\}$
(so the input alphabet of $M'$ consists 
of counter decreasing names).
Given an input $x \in \Delta_{(k,d)}^*$, $M'$ simulates
the computation of $M$ on some input $w \in \Sigma^*$
by guessing the symbols comprising $w$ and checking that
the sequence of counter decreases during the
accepting computation is $x$.  $M'$ accepts if and
only if $M$ accepts.
Clearly, $M$ is in $\NCM(\LB_d)$ if and only if $L(M')  \subseteq
\Delta_{(k,d)}^*$ is letter-bounded which is decidable
by Lemma \ref{1-bounded-1}.

The proof for the case of $\LB_i$ is similar.
%
%
For this case the input alphabet of $M'$ would be $\Delta_{(k,c)}
= \{C_1, \ldots, C_k \}$. Here $M'$, when
given an input $x \in \Delta_{(k,c)}^*$, simulates 
the computation of $M$ on some input $w \in \Sigma^*$
and checks that the sequence of counter increases
during the accepting computation is $x$. 

For the case of $\LB$, construct from $M$ a $k$-counter
well-formed $\NCM$ $M'$ which, when given an
input $x \in \Delta_k^*$, simulates $M$ on some input
$w \in \Sigma^*$ (by guessing
the symbols comprising $w$) and checking that the
sequence of counter increases and decreases in the
accepting computation is $x$.
Then, $M$ is in $\LB$ if and only if $L(M') \subseteq
\Delta_k^*$ is letter-bounded which is decidable.

For $\NCM(\StLB))$, the decision
procedure for $\LB$ is followed to determine the
$a_1, \ldots, a_k$ such that $L(M) \subseteq a_1^* \cdots a_k^*$,
then verify the stratified property.

Finally, the case of $\LB_i\LB_d$ is similar to $\LB$,
but determine that
the counter behavior of $M$ during any accepting
computation consists of increasing counters followed
by decreasing counters.
\qed
\end{proof}

We are not able to show decidability or undecidability for the cases
of $\BD_i\LB_d$,  $\LB_i\BD_d$, and $\BD$ at this
time, and these problems remain open.  However, two somewhat weaker
results are proven.

Let $M$ be an $\NPCM$ over an input alphabet
$\Sigma$. For $m \geq 1$, define a new alphabet
$\Sigma_m = \{ [x] ~|~ x \in \Sigma^+, |x| = m \}$.
Construct an $\NPCM$ $M_m$ over input alphabet $\Sigma_m$
which, on input $w' = [x_1] \cdots [x_r]$
(each $[x_i] \in \Sigma_m$) simulates the
computation of $M$ on input $w = x_1 \cdots x_r$.
Thus, $M_m$ reads $w'$ from left-to-right and
for each symbol $[x_i]$, it simulates the computation
of $M$ on $x_i$ before reading the next symbol $[x_{i+1}]$.
$M_m$ accepts  $w'$ if and only if $M$ accepts $w$. 
%

Clearly, from the construction of $M_m$,
if $w' = [x_1] \cdots [x_r]$ is in $L(M_m)$, then
$x_1 \cdots x_r$ is $L(M)$.
Conversely, if $w$ is in $L(M)$ and $|w|$ is divisible
by $m$, then $[x_1] \cdots [x_r]$ in $L(M_m)$,
where $x_1 \cdots x_r = w$.

The notation above will be used in Lemmas  \ref{1-bounded-2}
and \ref{decide1}.

\begin{lemma} \label{1-bounded-2}
Let $M$ be an $\NPCM$ such that the lengths of
all strings in $L(M)$ are divisible by $m$.
Then $M$ is $m$-bounded if and only if 
$L(M_m)$ is 1-bounded.
\end{lemma}
\begin{proof}
Suppose $L(M_m)$ is 1-bounded, i.e., there
are symbols $[x_1], \ldots, [x_k] \in \Sigma_m$
such that $L(M_m) \subseteq [x_1]^* \cdots [x_k]^*$.
Let $w$ be any string in $L(M)$.  By assumption, $|w|$ is
divisible by $m$. 
Hence, $w = y_1 \cdots y_s$, where $|y_i| = m$ for each $i$.
Clearly, by the construction of $M_m$, 
$w' = [y_1] \cdots [y_s]$ is in $L(M_m)$.  Further,
$w' \in [x_1]^* \cdots [x_k]^*$.  It follows
that $w$ is in $x_1^* \cdots x_k^*$ and,
therefore, $L(M) \subseteq x_1^* \cdots x_k^*$.
This shows that $L(M)$ is $m$-bounded. 
 
Conversely, suppose $M$ is $m$-bounded.  Then there are
strings $x_1, \ldots, x_k \in \Sigma^+$, $|x_i| =m$
(for each $i$) such that
$L(M) \subseteq x_1^* \cdots  x_k^*$.    
Hence, each string $w$ in $L(M)$ is of the form
$x_1^{i_1} \cdots x_k^{i_k}$ for some $i_1, \ldots, i_k$.

It will be shown that $L(M_m)$ is a subset of $[x_1]^* \cdots [x_k]^*$.
Suppose $w' = [y_1]  \cdots [y_s]$ is in $L(M_m)$.
Then $w = y_1 \cdots y_s$ is in $L(M)$.  Hence,
$w = y_1 \cdots y_s  =  x_1^{i_1} \cdots x_k^{i_k}$.
Since $|y_1| = \cdots = |y_s| = |x_1| = \cdots = |x_k| = m$,
it follows that $s = i_1 + \cdots + i_k$,
and $w' = [y_1] \cdots [y_s] =  [x_1]^{i_1} \cdots [x_k]^{i_k}$.
Hence, $L(M_m) \subseteq [x_1]^* \cdots [x_k]^*$.
\qed
\end{proof}

\begin{lemma} \label{decide1}
It is decidable, given $m \ge 1$ and an $\NPCM$ $M$,
whether $L(M)$ is $m$-bounded.
Moreover, if it is $m$-bounded,
we can effectively find a $k$ and not-necessarily
distinct words $w_1, \ldots, w_k$, $|w_i| = m$ (for each $i$)
such that $L(M) \subseteq  w_1^* \cdots w_k^*$.
\end{lemma}
\begin{proof}
First construct from $M$ another $\NPCM$ $M_0$
which has one additional counter $C_0$ than $M$.  
On a given input $w$, $M_0$ simulates $M$ while
storing $|w|$ in $C_0$.  When $M$ accepts,
$M_0$ checks (by decrementing $C_0$) and accepts
$w$ if the number in $C_0$ is not divisible by $m$.
If $L(M_0) \ne \emptyset$ (which is decidable
by Proposition \ref{known} part 1), then not all words are divisible by $m$ (hence $M$
is not $m$-bounded).

If $L(M_0) = \emptyset$, construct from $M$ the $\NPCM$ $M_m$.
We can determine if $L(M_m)$ is 1-bounded (by Lemma \ref{1-bounded-1}).
Furthermore, $L(M)$ is $m$-bounded if and only if
$L(M_m)$ is 1-bounded, 
by Lemma \ref{1-bounded-2}. Hence, we can determine if $L(M)$ is $m$-bounded.

Finally, if $L(M_m)$ is 1-bounded, 
we can effectively find (by exhaustive search using
Proposition  \ref{known} part 4), a $k\ge 1$ and not-necessarily
distinct symbols $[x_1], \ldots, [x_k] \in \Sigma_m$
such that $L(M_m) \subseteq [x_1]^* \cdots [x_k]^*$.
Then $L(M) \subseteq x_1^* \cdots x_k^*$.
\qed
\end{proof}

This can be used as follows:
\begin{proposition}
\label{boundeddecidable}
It is decidable, given a well-formed $k$-counter $\NCM$ machine $M$, 
and $n \ge 1$,
whether $M \in \NCM(\BD)$ where the bounded language
specification $w_1^* \cdots w_m^* $  of the counter
behavior is such that $|w_1| + \cdots + |w_m| \le n$.
\end{proposition}
\begin{proof}

First, as in Proposition \ref{fixedI}, construct
from $M$ another $k$-counter $\NCM$ $M'$ whose input
alphabet is $\Delta_k = \{C_1, D_1, \ldots, C_k, D_k\}$.
$M'$ on input $x \in \Delta_k^*$ simulates the
computation  $M$ on some input $w$ (guessing the symbols
comprising $w$) and checking that the counter behavior
when $M$ accepts is $x$.

Now for a given $n$, there are only a finite number of
desired counter behaviors possible.  Thus, all are
systematically enumerated.  For each such
behavior, $P$, construct an $\DFA$ $M_P$ accepting $P$,

Next construct a $\DFA$ $M''$ which accepts
the union of all the $P$'s.   Clearly
$M$ is in $\BD$ (with respecting the bound $n$)
if and only if $L(M') \subseteq L(M'')$, which is decidable,
since the containment of an $\NCM(k,1)$ language
in a regular set is decidable \cite{Ibarra1978}.
\qed
\end{proof}


%
%

Finally, we briefly look at the question of deciding
whether a given language $L$ in a family 
of languages $\LL$ is contained in a proper subset
$\cal S$ of $\LL$.
We will need the following result called Greibach's Theorem \cite{Greibach}:
\begin{proposition} \label{Gthm} \cite{Greibach}
(Greibach)
Let $\LL$ be any family of languages
that contains all regular languages, is effectively
closed under union and
concatenation with regular languages, and
for which the question ``Is $L = \Sigma^*$?''
is undecidable for any sufficiently large alphabet $\Sigma$.
Let $\cal S$ be any proper subset of $\LL$ that contains
all regular languages
and is closed under right quotient by any symbol.
Then it is undecidable, given an
an arbitrary language $L$ in $\LL$, whether $L$ is in $\cal S$.
\end{proposition}

With this, consider the following:
\begin{proposition} \label{universe}
For each ${\cal I}$ in Definition \ref{def:families}, it is undecidable
whether $L \in \LL(\NCM({\cal I}))$ is equal to $\Sigma^*$.
\end{proposition}
\begin{proof}
It is sufficient to prove for $\LB_i\LB_d$ and $\StLB$. This follows from the proof of undecidability for $\NCM(1,1)$ in \cite{Baker1974}, which accepts the language of invalid computations of a Turing machine using a single-reversal of a single counter, with instructions in $C_1^* D_1^*$.
\qed
\end{proof}

From Greibach's Theorem:
\begin{proposition} The following are true:
\begin{enumerate}
\item Let $\LL$ be any of the following: $\LL(\NCM)$,
$\LL(\NCM(\LB_d))$, $\LL(\NCM(\LB_i))$, $\LL(\NCM(\BD))$,
and let $\cal S$ = $\LL(\NCM(\LB))$.
Then it is undecidable,
given $L$ in $\LL$, whether $L$ is in $\cal S$.
\item Let $\LL$ be  any of the
following:
$\LL(\NCM)$,
$\LL(\NCM(\LB_d))$, $\LL(\NCM(\LB_i))$, $\LL(\NCM(\BD))$,
$\LL(\NCM(\LB))$, and let ${\cal S}$ be any of the following families: $\LL(\NCM(\LB_i\LB_d))$,
$\LL(\NCM(\StLB))$.
Then it is is undecidable, given $L$ in $\LL$, whether
$L$ is in $\cal S$.
\end{enumerate}
\end{proposition}
\begin{proof}
For 1), $\LL$ and $\cal S$ satisfy the hypothesis 
of Proposition \ref{Gthm} by Proposition \ref{universe} and by noting that  $\cal S$ is a proper subset of $\LL$,
since from Proposition \ref{witness}, $\{a^n b^n c^n  ~|~  n > 0 \}$
is not in $\cal S$ but is clearly in $\LL$.

For 2), this follows from Proposition \ref{Gthm}
and the fact (see Corollary \ref{cor6}) that
$\{a^i b^i c^j d^j  ~|~ i, j \ge 1 \}$
is in $\LL(\NCM(\LB)$) but not in $\LL(\NCM(\LB_i\LB_d))$,
and $\{a^i b^j a^i b^j ~|~ i,j \ge 1\}$ 
is in $\LL(\NCM(\LB)$) but not in $\LL(\NCM(\StLB))$.
\qed
\end{proof}

In the previous section, we conjectured that:
\begin{enumerate}
\item
$\LL(\NCM(\BD))$ is a proper subset of 
$\LL(\NCM(\LB_d))$ (resp., $\LL(\NCM(\LB_i))$.
\item
$\LL(\NCM(\LB_d))$ (resp., $\LL(\NCM(\LB_i))$
is a proper subset of $\LL(\NCM)$.
\end{enumerate}

\noindent
If the conjectures are correct, then by Greibach's Theorem,
it would follow that:
\begin{enumerate}
\item
It is undecidable, given $L$ in  $\LL(\NCM(\LB_d))$
(resp., $\LL(\NCM(\LB_i))$, whether $L$ is in $\LL(\NCM(\BD_i))$.
\item
It is undecidable, given $L$ in  $\LL(\NCM)$, whether
$L$ is in $\LL(\NCM(\LB_d))$ (resp., $\LL(\NCM(\LB_i))$.
\end{enumerate}

\section{Weakly Satisfying Instruction Languages}
\label{weakproperties}

In Section \ref{sec:instruction}, instruction languages were introduced, along with machines satisfying those
instruction languages. In particular, a well-formed $k$-counter machine $M$ was said to satisfy
instruction language $I \subseteq \Delta_k^*$ if every sequence of transitions $\alpha \in T^*$ used in an accepting
computation has $h_{\Delta}(\alpha) \in I$. A weaker model is also quite natural in this case, whereby instead of requiring
that every accepting computation give a word in the instruction language, at least one accepting computation $\alpha$ 
of every word in $L(M)$ needs to have $h_{\Delta}(\alpha) \in I$.
Formally, it is said that $M$ {\em weakly satisfies} instruction language $I \subseteq \Delta_k^*$ if, 
for every word $w \in L(M)$, there exists a derivation, $(q_0, w\lhd,0,\ldots, 0) \vdash_M^{\alpha} (q,\lambda, c_1, \ldots, c_k), q\in F$, such
that $h_{\Delta}(\alpha) \in I$. Given a family of languages ${\cal I}$ over $\Delta_k$, let $\NCM_w(k,{\cal I})$ be the subset of
well-formed $k$-counter machines that weakly satisfy $I$ for some $I \in {\cal I}$. Then $\NCM_w(k,{\cal I})$ is called
the $k$-counter weakly satisfied ${\cal I}$-instruction machines (and $\LL(\NCM_w(k,{\cal I}))$ for the languages), and $\NCM_w({\cal I})$ is the weakly
satisfied ${\cal I}$-instruction machines (and $\LL(\NCM_w({\cal I}))$ are the languages) where all $k$-counter machines in $\NCM_w({\cal I})$ weakly satisfy $I$, for
some $I \in {\cal I}$ over $\Delta_k$. 

First, note the following regarding $1$-counter machines.
\begin{remark}
Every well-formed $1$-counter machine $M$ satisfies, and weakly satisfies, $C_1^* D_1^*$.
\end{remark}
This is because every accepting computation for a well-formed $1$-counter machine satisfies this instruction language.
Therefore, it is trivially decidable to test whether a $1$-counter machine satisfies, or weakly satisfies this language.
\begin{proposition}
Let ${\cal I} \in \{\LB_i\LB_d, \StLB, \LB, \LB_d, \LB_i, \LB_{\cup}\}$. It is decidable, given a well-formed $1$-counter machine $M$, whether $M$ weakly satisfies $I \in {\cal I}$.
\end{proposition}
\begin{proof}
Every $I \subseteq \{C_1, D_1\}^*$ in every such ${\cal I}$ above, intersected with $C_1^* D_1^*$ (every accepting computation
of a well-formed $1$-counter machine must use only instruction sequences of this form) is equal to $C_1^* D_1^*$. Therefore, by the remark above, it is decidable.
\qed
\end{proof}
Several instruction language families are noticeably absent from the 
proposition above, which we will consider next. In some sense, this decidability is only for trivial reasons, as we see that if just remove the empty word from the instruction language, the problem becomes undecidable.
\begin{proposition}
It is undecidable, given a well-formed $1$-counter machine $M$, whether
$M$ weakly satisfies $I = C_1^+ D_1^+$ (resp. $C_1^* (D_1 D_1)^*$, and
$(C_1C_1)^* D_1^*$).
\end{proposition}
\begin{proof}
First, consider $I = C_1^+ D_1^+$. 
Given an arbitrary $\LL(\NCM(1,1))$ language $L\subseteq \Sigma^*$, 
there exists
a well-formed $1$-counter machine $M$ accepting $L$ that 
always increases the counter (and decreases it).

Create a new well-formed $1$-counter machine $M'$ that operates as follows, by doing one of the following two things nondeterministically:
\begin{enumerate}
\item $M'$ simulates $M$, then if $M$ accepts, $M'$ does as well.
\item $M'$ reads the input without using the counter at all, and accepts.
\end{enumerate}
Clearly, $M'$ satisfies $I \cup \{\lambda\}$ and $L(M')= \Sigma^*$.
But $M'$ weakly satisfies $I$ if and only if $L(M) = \Sigma^*$.
But this is a known undecidable problem for $\NCM(1,1)$ \cite{Baker1974}.

For the instruction language $I = (C_1C_1)^*D_1^*$, the machine
$M$ is assumed without loss of generality to always increase
and decrease the counter an even number of times in every accepting
computation, and the machine $M'$ need only be modified to, in step 2, have $M'$
read the input, add $1$ to the counter, decrease the counter,
then accept (so the instruction sequence is $C_1D_1$).
Then $M'$ weakly satisfies $I$ if and only if $L(M) = \Sigma^*$.
Similarly with $C_1^* (D_1D_1)^*$.
\qed
\end{proof}
\begin{corollary}
For each ${\cal I} \in \{\BD_i\LB_d,\LB_i\BD_d,\BD\}$, the following
problem is undecidable: given a well-formed $1$-counter machine $M$, and a fixed language $I \in {\cal I}$, 
whether $M$ weakly satisfies $I$.
\end{corollary}
This follows since $(C_1C_1)^*D_1^*$ is in $\BD_i\LB_d$ and $\BD$,
and $C_1^* (D_1D_1)^*$ is in $\LB_i\BD_d$.

Furthermore, if we allow two counters, then this problem becomes
undecidable for all of the language families of
Definition \ref{def:families}.
\begin{proposition} 
Let ${\cal I}$ be any instruction language family from
Definition \ref{def:families}. 
It is undecidable, given a well-formed $2$-counter $\NCM$ machine 
$M$, and a fixed language $I\in {\cal I}$, whether $M$ weakly
satisfies instruction language $I$.
\end{proposition}
\begin{proof}
Consider the fixed $I = C_2^* C_1^* D_1^* D_2^*$.
Let $L \subseteq \Sigma^*$ be an 
$\NCM(1,1)$ language. Then, there is a well-formed
$1$-counter $\NCM$ machine $M$ accepting $L$.  
Let $V = \{c_1, c_2, d_1, d_2\}$ be  a
new set of symbols. Construct a well-formed $2$-counter
$\NCM$ machine $M'$ over alphabet $\Sigma \cup V$,
which operates as follows when given an input $w$, by doing one of the following
two things nondeterministically:
\begin{enumerate}
\item On a $\lambda$ move, $M'$ increments counter $2$ to one. Then
$M'$ makes sure that the input 
$w = xy$, where $x \in \Sigma^*, y \in V^*$,
and simulates $M$ on $x$ using counter $1$.  
If $M$ accepts $x$, $M'$ decrements
counter $2$ and then verifies that $y \in V^*$ and then accepts.

\item 
$M'$ makes sure that the input $w = xy$, where $x \in \Sigma^*, 
y \in V^*$,
and reads past $x$.  Then when reading segment $y$, whenever 
it sees $c_1$
(resp., $c_2$) it increments counter $1$ (resp. counter $2$), 
and whenever it sees
$d_1$ (resp. $d_2$), it decrements counter $1$ 
(resp. counter $2$) provided
it is positive. Moreover $M'$ checks that all increments of counter
$1$ (resp. counter $2$) occur before any decrement of counter $1$
(resp. counter $2$). $M'$ accepts $w =xy$ when at the end of $y$ the counters are zero.
\end{enumerate}

Computations of type 1 satisfy $C_2 C_1^* D_1^* D_2$, and type 2 satisfy $X = \{z  \mid  z \in \{C_1, D_1, C_2, D_2\}^*,$  all 
$c_1$'s occur before any $d_1$,
all $c_2$'s occur before any $d_2$, $|z|_{C_1} = |z|_{D_1},  |z|_{C_2} = |z|_{D_2} \}$.
Clearly $M'$ accepts $L' = L V^* \cup \Sigma^* X'$, where $X'$ is obtained from $X$ by making each letter lowercase.  
If $L(M) = \Sigma^*$, then $L' = \Sigma^* V^* \cup \Sigma^* X'$, and since $\Sigma^* X' \subseteq \Sigma^* V^*$,
$M'$ weakly satisfies $I$.
If $L(M) \neq \Sigma^*$, then say $x \in \Sigma^* - L(M)$ and let $y \in X'$ where $y$ has counter
behavior not in $I$; then $xy \in L'$ and does not have an accepting computation with counter behavior
in $I$.

It follows that $M$ weakly satisfies $I$ if
and only if 
$L(M) = \Sigma^*$,
which is undecidable \cite{Baker1974}.
\qed
\end{proof}

Next, the language family accepted by $\NCM$
machines satisfying some instruction language family
${\cal I}$ will be compared to those weakly satisfying the same family.
\begin{proposition}
\label{satifyweaklysatisfy}
Let $I \in \LL(\NCM)$ be an instruction language, and let
$M$ be a well-formed $\NCM$ machine weakly satisfying $I$. Then there exists a well-formed $\NCM$ machine $M'$ satisfying $I$
such that $L(M) = L(M')$.
\end{proposition}
\begin{proof}
Create $M'$ to simulate $M$, but it only uses sequences
of instructions that are in $I$. That is, it only accepts if
the simulation of $M$ accepts, and the sequence
of instructions used in the derivation is in $I$, which
can be verified in parallel using additional counters.

It is trivial that $L(M') \subseteq L(M)$. The reverse inclusion
can be seen since, any word $w \in L(M)$ has some derivation
with instructions in $I$. Thus, there is a computation in $M'$ that
accepts $w$ as well. Hence, $L(M') = L(M)$.

And $M'$ satisfies $I$ since every accepting computation of a 
word $w \in L(M')$ has a derivation in $I$ by the construction 
of $M'$.
\qed
\end{proof}
It should be noted that every instruction language family
considered so far, such as those in Definition \ref{def:families},
are all regular, and so the restriction $I \in \LL(\NCM)$ applies to
all families in this paper.

\begin{proposition}
Let ${\cal I}$ be a family of instruction languages, such that
${\cal I} \subseteq \LL(\NCM)$. Then, 
$\LL(\NCM({\cal I})) = \LL(\NCM_w({\cal I}))$.
\end{proposition}
\begin{proof}
It is immediate that every well-formed counter machine $M$ 
that satisfies $I \in {\cal I}$, also weakly satisfies
$I$. The other direction follows from
Proposition \ref{satifyweaklysatisfy}.
\qed
\end{proof}

\section{Conclusions and Future Directions}

In this paper, restrictions of the well-studied one-way reversal-bounded multicounter machines are studied. This is considered through the use of instruction languages, which restricts the allowable patterns of counter addition and subtraction behaviors. By considering some simple regular patterns, it is possible to e.g.\ accept exactly the smallest full trio containing the bounded context-free languages, or the smallest full trio containing the bounded semilinear languages. These characterizations provide several easy methods for studying the bounded languages in any semilinear full trio of interest. Indeed, they are applied here to show easy proofs that the bounded languages in seemingly quite different models coincide ($\NCM$, a restriction on Turing machines, multi-pushdown machines). Therefore, instruction languages are a useful tool for comparing families of languages.

Furthermore, the computational power of the $\NCM$ restrictions are compared in terms of the specific families of instruction languages. While some separation results are provided, several remain open. For example, are there any languages that can be accepted where counter decreases satisfy a letter-bounded pattern that cannot be accepted by a machine where both counter increases and decreases together satisfy a bounded pattern?

Decidability properties are also given to decide whether a given counter machine satisfies some instruction 
language in a given instruction language family. For some instruction language families, this is decidable, whereas the problem remains open for others considered. However, if the specific instruction language itself is given rather than only its family, then this problem is always decidable when the instruction language are from any family considered here.

\section*{Acknowledgements}
We thank an anonymous reviewer whose suggestions improved the readability of the paper, as well as the presentation of some of the proofs.

\bibliography{bounded}{}

\begin{thebibliography}{10}
\expandafter\ifx\csname url\endcsname\relax
  \def\url#1{\texttt{#1}}\fi
\expandafter\ifx\csname urlprefix\endcsname\relax\def\urlprefix{URL }\fi
\expandafter\ifx\csname href\endcsname\relax
  \def\href#1#2{#2} \def\path#1{#1}\fi

\bibitem{DLT2016}
O.~Ibarra, I.~McQuillan, On families of full trios containing counter machine
  languages, in: S.~Brlek, C.~Reutenauer (Eds.), Lecture Notes in Computer
  Science, Vol. 9840 of 20th International Conference on Developments in
  Language Theory, DLT 2016, 2016, pp. 216--228.

\bibitem{HU}
J.~E. Hopcroft, J.~D. Ullman, Introduction to Automata Theory, Languages, and
  Computation, Addison-Wesley, Reading, MA, 1979.

\bibitem{GiSp68a}
S.~Ginsburg, E.~H. Spanier, Derivation-bounded languages, Journal of Computer
  and System Sciences 2 (1968) 118--250.

\bibitem{CheckingStack}
S.~Greibach, Checking automata and one-way stack languages, Journal of Computer
  and System Sciences 3~(2) (1969) 196--217.

\bibitem{checkingstacksTCS}
O.~Ibarra, I.~McQuillan, Variations of checking stack automata: Obtaining
  unexpected decidability properties, Theoretical Computer Science 738 (2018)
  1--12.

\bibitem{Ibarra1978}
O.~H. Ibarra, Reversal-bounded multicounter machines and their decision
  problems, Journal of the ACM 25~(1) (1978) 116--133.

\bibitem{StoreLanguages}
O.~Ibarra, I.~McQuillan, On store languages of languages acceptors, Theoretical
  Computer Science 745 (2018) 114--132.

\bibitem{Baker1974}
B.~S. Baker, R.~V. Book, Reversal-bounded multipushdown machines, Journal of
  Computer and System Sciences 8~(3) (1974) 315--332.

\bibitem{DCMDeletion}
J.~Eremondi, O.~H. Ibarra, I.~McQuillan, Deletion operations on deterministic
  families of automata, Information and Computation 256 (2017) 237--252.

\bibitem{DCMInsertion}
J.~Eremondi, O.~Ibarra, I.~McQuillan, Insertion operations on deterministic
  reversal-bounded counter machines, Journal of Computer and System Sciences
  104 (2019) 244--257.

\bibitem{HagueLin2011}
M.~Hague, A.~Lin, Model checking recursive programs with numeric data types,
  in: G.~Gopalakrishnan, S.~Qadeer (Eds.), Computer Aided Verification, Vol.
  6806 of Lecture Notes in Computer Science, Springer Berlin Heidelberg, 2011,
  pp. 743--759.

\bibitem{IBARRA2002165}
O.~Ibarra, J.~Su, Z.~Dang, T.~Bultan, R.~Kemmerer, Counter machines and
  verification problems, Theoretical Computer Science 289~(1) (2002) 165--189.

\bibitem{Dang2001}
Z.~Dang, Binary reachability analysis of pushdown timed automata with dense
  clocks, in: G.~Berry, H.~Comon, A.~Finkel (Eds.), Computer Aided
  Verification: 13th International Conference, CAV 2001, Proceedings, 2001, pp.
  506--517.

\bibitem{IbarraDang}
G.~Xie, Z.~Dang, O.~Ibarra, A solvable class of quadratic diophantine equations
  with applications to verification of infinite-state systems, in: J.~Baeten,
  J.~Lenstra, J.~Parrow, G.~Woeginger (Eds.), Automata, Languages and
  Programming: 30th International Colloquium, ICALP 2003, Proceedings, 2003,
  pp. 668--680.

\bibitem{Gurari1981220}
E.~M. Gurari, O.~H. Ibarra, The complexity of decision problems for finite-turn
  multicounter machines, Journal of Computer and System Sciences 22~(2) (1981)
  220--229.

\bibitem{IbarraSeki}
O.~H. Ibarra, S.~Seki, Characterizations of bounded semilinear languages by
  one-way and two-way deterministic machines, International Journal of
  Foundations of Computer Science 23~(6) (2012) 1291--1306.

\bibitem{CIAA2016}
O.~Ibarra, I.~McQuillan, On bounded semilinear languages, counter machines, and
  finite-index {ET0L}, in: Y.-S. Han, K.~Salomaa (Eds.), Lecture Notes in
  Computer Science, Vol. 9705 of 21st International Conference on
  Implementation and Application of Automata, CIAA 2016, Seoul, South Korea,
  2016, pp. 138--149.

\bibitem{multipushdown}
L.~Breveglieri, A.~Cherubini, C.~Citrini, S.~Reghizzi, Multi-push-down
  languages and grammars, International Journal of Foundations of Computer
  Science 7~(3) (1996) 253--291.

\bibitem{GGr1}
S.~G. Seymour~Ginsburg, Abstract families of languages, in: J.~H.
  Seymour~Ginsburg, Sheila~Greibach (Ed.), Studies in Abstract Families of
  Languages, Vol.~87 of Memoirs of the American Mathematical Society, American
  Mathematical Society, 1969, pp. 1--32.

\bibitem{G75}
S.~Ginsburg, Algebraic and Automata-Theoretic Properties of Formal Languages,
  North-Holland Publishing Company, Amsterdam, 1975.

\bibitem{G78}
S.~Greibach, Remarks on blind and partially blind one-way multicounter
  machines, Theoretical Computer Science 7 (1978) 311--324.

\bibitem{smallestBounded}
J.~Kortelainen, T.~Salmi, There does not exist a minimal full trio with respect
  to bounded context-free languages, in: G.~Mauri, A.~Leporati (Eds.), Lecture
  Notes in Computer Science, Vol. 6795 of 15th International Conference on
  Developments in Language Theory, DLT 2011, Milan, Italy, 2011, pp. 312--323.

\bibitem{GinsburgCFLs}
S.~Ginsburg, The Mathematical Theory of Context-Free Languages, McGraw-Hill,
  Inc., New York, NY, USA, 1966.

\bibitem{IbarraRavikumar}
O.~Ibarra, B.~Ravikumar, On bounded languages and reversal-bounded automata,
  Information and Computation 246~(C) (2016) 30--42.

\bibitem{ResetMachines}
R.~V. Book, S.~A. Greibach, C.~Wrathall, Reset machines, Journal of Computer
  and System Sciences 19~(3) (1979) 256--276.

\bibitem{Harju2002278}
T.~Harju, O.~Ibarra, J.~Karhumäki, A.~Salomaa, Some decision problems
  concerning semilinearity and commutation, Journal of Computer and System
  Sciences 65~(2) (2002) 278--294.

\bibitem{boundedAFLs}
P.~Turakainen, On some bounded semi{AFL}s and {AFL}s, Information Sciences
  23~(1) (1981) 31--48.

\bibitem{DLT2015journal}
J.~Eremondi, O.~Ibarra, I.~McQuillan, On the density of context-free and
  counter languages, International Journal of Foundations of Computer Science
  29~(2) (2018) 233--250.

\bibitem{Greibach}
S.~Greibach, A note on undecidable properties of formal languages, Mathematics
  Systems Theory 2~(1) (1968) 1--6.

\end{thebibliography}
\bibliographystyle{elsarticle-num}

\end{document}